\documentclass[letterpaper,9pt]{article}
\usepackage{graphicx}
\usepackage{caption}
\usepackage{subcaption}
\usepackage{multirow}
\usepackage{url}
\usepackage{array}
\usepackage{cite}
\usepackage{color}
\usepackage{arydshln}
\usepackage[margin=1in]{geometry}
\usepackage{authblk}
\usepackage{chngpage}
\usepackage{hyperref}
\usepackage{booktabs}

\begin{document}

\title{NFV Platform Design: A Survey}

\author[]{Tianzhu~Zhang \\ Email: \href{mailto:tianzhu.zhang1989@gmail.com}{tianzhu.zhang1989@gmail.com} }
\date{}  
\maketitle 

\begin{abstract}
Due to the inherently inefficient service provisioning in traditional networks, Network Function Virtualization (NFV) continues to attract attention from both industry and academia.
By replacing the purpose-built, expensive, proprietary network equipment with software network functions consolidated on commodity hardware, NFV envisions a shift towards a more agile and open service provisioning paradigm with much lower capital expenditure (CapEx) and operational expenditure (OpEx).
Nonetheless, like any complex system, NFV platforms typically comprise numerous software and hardware components and often incorporate disparate design choices driven by distinct motivations or use cases.
This broad collection of convoluted alternatives makes it extremely arduous for network operators to make proper choices. 
Although numerous efforts have investigated various aspects of NFV, none have specifically focused on NFV platforms or explored the design space.
In this paper, we present a comprehensive survey on NFV platform design. Our study solely targets existing NFV platform implementations.
We begin with an architectural view of the standard reference NFV platform and present our taxonomy of existing NFV platforms by their principal design purpose. We then thoroughly explore the design space and elaborate on the implementation choices adopted by each platform.
We believe that our study provides a detailed guideline for network operators or service providers to select or implement the most appropriate NFV platforms based on their requirements. 
\footnote{This document originally served as a complementary document for a published IEEE TNSM paper~\cite{9302614}. It will be updated periodically to include the latest NFV systems and design choices. }
\end{abstract}


\section{Introduction} \label{sec:introduction}
Traditionally, network services are provisioned using purpose-built, proprietary hardware appliances (or middleboxes). Middleboxes encompass a wide range of specialized functions for forwarding, classifying, or transforming traffic based on packet content.
Examples of middleboxes include, but are not limited to, L2 Switching, Routing, Network Address Translation (NAT), Firewall (FW), Deep Packet Inspection~(DPI), Intrusion Detection System (IDS), Load Balancer~(LB), WAN optimizer, and stateful proxy. Nowadays, middleboxes are ubiquitous in enterprise networks~\cite{sherry2012making}.
However, with increasingly diverse user requirements and the rapid growth of Internet traffic, in terms of both volume and heterogeneity~\cite{cisco2018cisco}, hardware middleboxes are beginning to exhibit several fundamental disadvantages. First off, middleboxes are generally expensive to acquire and typically require domain-specific knowledge to manage, resulting in large capital expenditure (CapEx) and operational expenditure (OpEx). Also, adding customized functionality is extremely time-consuming, if not impossible, and it sometimes takes an entire purchase cycle (e.g., four years) to bring in equipment with new features~\cite{martins2014clickos}. Such tight coupling with the hardware production cycle considerably hampers network innovation and prolongs time-to-market.
Deploying new network services (NSs) is also a tedious process, as technicians must visit specific sites and place middleboxes in a pre-defined order to form the correct service function chains (SFCs). Service instantiation might even take days. Worse still, service maintenance usually involves constant repetition of the same process.
Furthermore, due to inherent inflexibility, it is nontrivial for hardware middleboxes to elastically scale in and out in response to shifting demand or other system dynamics. Consequently, network operators usually resort to peak-load provisioning, which in turn leads to ineffective resource utilization and extravagant energy consumption.

\begin{figure}[!tb]
 \centering
  \includegraphics[width=0.7\textwidth]{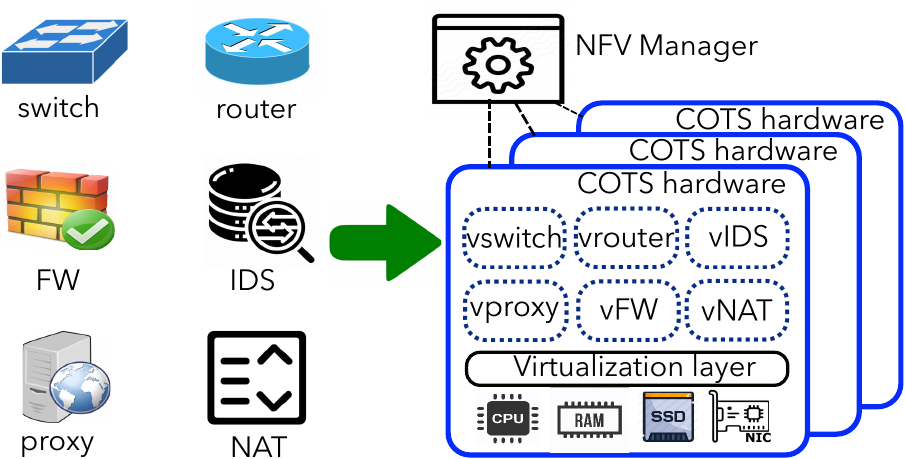}
   \caption{Traditional vs. NFV paradigm}
 \label{fig:compare}
 \vspace{-0.3cm}
\end{figure}

To improve service provisioning and eliminate network ossification, telecommunication operators have begun to pursue new solutions that guarantee both cost-effectiveness and flexibility.
The advent of Software Defined Networking (SDN)~\cite{mckeown2008openflow} and Network Function Virtualization (NFV)~\cite{mijumbi2015network} provides alternative approaches for network management and service provisioning. 
SDN decouples the control plane from the data plane and leverages a logically centralized controller to configure programmable switches based on a global view, while NFV replaces specialized middleboxes with software-based Virtual Network Functions (VNFs) consolidated on Commodity Off-the-Shelf (COTS) hardware.
The key to their success lies in separating the evolution timelines of software network functions and specialized hardware, thereby fully unleashing the potential of the former.
An illustrative example contrasting the NFV paradigm with the traditional network is shown in Fig.~\ref{fig:compare}.
Compared to the traditional service provisioning paradigm based on hardware middleboxes, NFV achieves cost-effectiveness by consolidating multiple instances of VNFs on high-volume yet inexpensive servers, routers, or storage.
Service provisioning in NFV is also highly simplified as the previously troublesome tasks, such as middlebox deployment, monitoring, migration, and scaling, can be optimally automated through software control mechanisms.
It is thus convenient for NFV solutions to exploit available resources and management tools of the cloud infrastructure or network edge.
Also, NFV significantly promotes network innovation and accelerates time-to-market by reducing the time required to develop network functions to writing software programs using standard application programming interfaces (APIs).

Thanks to these indispensable merits, NFV continues to gain momentum in both industry and academia.
The first concerted effort towards NFV standardization began in 2012, with the appointment of the European Telecommunications Standards Institute (ETSI)~\cite{etsi} as Industry Specification Group. Currently, ETSI comprises more than 500 members worldwide, including major telecommunications operators, service providers, manufacturers, and universities. 
In the meantime, the continuous advancement of COTS hardware capabilities and the emergence of high-speed packet-processing techniques have significantly narrowed the previously large performance gap between software network functions and specialized middleboxes. Resources from other hardware components, such as Graphics Processing Units (GPUs), smartNICs, and in-path programmable network devices, can also be exploited to share the workload and alleviate the burden on the CPU. These technical impetuses immeasurably stimulate the growth of NFV.
In recent years, NFV has entered a new phase driven by the rapid adoption of cloud-native design principles in telecom infrastructure. From an operator's perspective, cloud-native NFV goes beyond merely replacing virtual machines with containers: it advocates microservice decomposition, declarative management, and automated lifecycle operations, with Kubernetes serving as the de facto orchestrator for the execution and management of containerized network functions (often referred to as Cloud Native Network Functions, CNFs). Industry bodies have also begun consolidating this trend into operator-facing principles and reference documents, which increasingly frame NFV as a broader \emph{Telco Cloud} capability rather than a VM-centric platform. \cite{ngmn_cloud_native_manifesto,ngmn_cloud_native_telco_platforms,etsi_wp65_telco_cloud}

In parallel with this cloud-native shift, the \emph{execution substrate} of NFV datapaths has also evolved rapidly since 2020.
First, in-kernel programmability (e.g., eBPF/XDP) has matured from a “fast packet hook” into a viable substrate for production-grade network functions and policies, motivating systems that focus on scalable deployment, configuration, and lifecycle management of in-kernel functions at fleet scale~\cite{benson2024netedit, YangEuroSys25eNetSTL}.
Second, heterogeneous offload targets are no longer limited to one-off accelerators: SmartNICs/DPUs, programmable switches, and FPGA-based gateways have become increasingly common building blocks for NFV platforms, reshaping the performance--flexibility tradeoffs and introducing new questions about resource pooling, multi-tenancy, and upgrade cadence~\cite{qiu2021clara, xing2023pipeleon, li2025nezha, pan2021sailfish, pan2024luoshen, lu2025albatross}.
Together, these trends broaden NFV from a VM-centric software dataplane into a heterogeneous, programmable substrate spanning host software, kernel fast paths, NIC-attached compute, and in-network devices.

Over the last decade, a large assemblage of NFV platforms has been deployed to spur innovation and evolution in NFV.
However, like any complex system, NFV platforms typically encompass numerous software and hardware components and embrace diverse design choices driven by their respective motivations or use cases. 
The design space of these platforms can be very expansive, ranging from high-level VNF development, such as VNF execution models, state management schemes, and API genres, to low-level infrastructure details, such as packet I/O frameworks, VNF interconnection methods, and virtualization techniques.
They also opt for various acceleration techniques, including compute batching, zero-copy packet transfer, parallelization, data prefetching, and computation offloading. Such a broad range of platform implementations, coupled with an even more extensive design space, makes it extremely difficult (if not impossible) for network operators to choose the most suitable solutions. The tradeoffs and caveats between different design choices are also unclear. 

Recent platform work further suggests that performance and scalability are increasingly shaped by \emph{operational} constraints, not only by dataplane micro-optimizations.
For example, elasticity requirements have driven attention to cold-start and instantiation overheads in cloud-style deployments (including microVM-style isolation and fast initialization techniques), which can dominate the service time budget under bursty traffic~\cite{agache2020firecracker, du2020catalyzer}.
At the same time, achieving high throughput on commodity servers now often relies on platform support for scaling and consolidation: rather than requiring developers to rewrite VNFs for concurrency, recent systems explore automatic or transparent multicore scaling and interference-aware consolidation~\cite{pereira2024maestro, yan2024transparent, li2023lemonnfv}.
For stateful services, another emerging direction is to reduce the coupling between \emph{state} and \emph{compute} so that scaling and recovery can be performed more flexibly, complementing classical state migration/redistribution mechanisms~\cite{bansal2023disaggregating}.
Several existing works have investigated some aspects of NFV, including VNF placement~\cite{li2016survey}, resource allocation~\cite{herrera2016resource}, service function chaining~\cite{bhamare2016survey}, and security~\cite{yang2016survey}, but none of them specifically focused on the design of NFV platforms, nor did they attempt to explore the design space or review different implementation choices. In \cite{de2019network,mijumbi2016management,yi2018comprehensive}, the authors investigated a subset of industrial NFV projects, which are complementary to our work.
Finally, as NFV platforms become more heterogeneous and multi-tenant, the community has placed renewed emphasis on \emph{trustworthiness} and \emph{operability}. Beyond traditional monitoring, recent work advocates platform interfaces and tooling for systematic diagnosis and performance isolation in consolidated deployments~\cite{iyer2022pix}.
Meanwhile, auditing and verification techniques are being revisited to better match operational reality, including settings where VNFs are provided by third parties or only binaries are available~\cite{liu2021auditbox, pirelli2022klint}.
These directions reinforce the motivation of our paper: the “right” NFV design increasingly depends on a coupled set of choices across programmability, orchestration, dataplane substrate, acceleration targets, and operational guarantees.

In this paper, we focus on existing NFV platforms and present a comprehensive survey of their design. The contribution of this paper can be summarized as follows:
\begin{itemize}
\item we classify existing NFV platforms by their primary purposes and review their internals.
\item we explore the NFV design space and discuss the various design choices adopted by existing platforms.
\end{itemize}

This paper is organized as follows: in Sec.~\ref{sec:architecture}, we give an architectural overview of the components of NFV platforms. We then present our taxonomy of existing platforms in Sec.~\ref{sec:taxonomy}. In Sec.~\ref{sec:design}, we propose a collection of critical design choices and survey the solutions adopted by different platforms. We draw the conclusion in Sec.~\ref{sec:conclusion}.


\section{NFV platform: an architectural overview} \label{sec:architecture}
\begin{figure*}[!tb]
 \centering
  \includegraphics[width=0.98\textwidth]{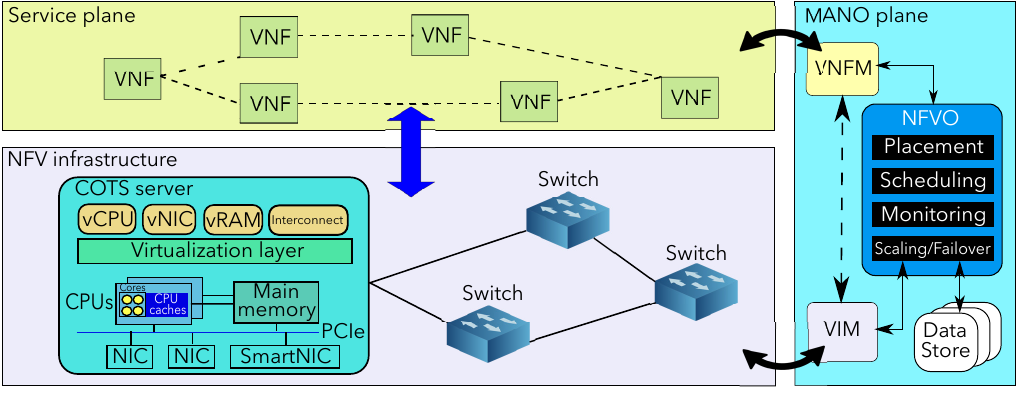}
 \caption{The architecture of a general NFV platform}
 \label{fig:arch}
\end{figure*}

We devote this section to providing an architectural overview of a typical NFV platform and reviewing its key components in depth. Although a reference architecture has been defined in the ETSI specification~\cite{etsi2013network}, most existing NFV platforms do not strictly follow it. As a result, we seek to combine the ETSI reference architecture with those of the existing platforms and present a generic view, as illustrated in Fig.~\ref{fig:arch}. An NFV platform generally consists of three primary components, namely the NFV Management and Orchestration (MANO) plane, the service plane, and the NFV Infrastructure (NFVI). The MANO plane provides centralized control over service provisioning and management. The NFVI comprises a collection of computational, storage, and network resources distributed across multiple infrastructure nodes. MANO plane components systematically monitor and schedule the resources to build a virtualized environment and accommodate different network services. The service plane contains a diverse collection of VNFs arranged in service chains to deliver the promised network services. These service chains are also carefully monitored and adjusted by the MANO plane components to efficiently multiplex the NFVI resources. In general, the service placement is enabled through coordinated operations between the MANO plane and the NFVI.

\subsection{MANO plane}
NFV Management and Orchestration (MANO) is the central point for service provisioning in NFV. A MANO system typically consists of three sub-systems: NFV Orchestrator (NFVO), Virtual Infrastructure Manager (VIM), and VNF Manager (VNFM). As shown in Fig.~\ref{fig:arch},
NFVO is responsible for the instantiation, management, and termination of network services. At present, an NFVO typically comprises multiple modules to perform different MANO operations. On the right part of Fig~\ref{fig:arch}, we illustrate four example modules. The placement module is responsible for determining the optimal deployment, possibly in an incremental fashion. When new services need to be deployed, the placement module analyzes the service descriptions or requirements specified by network operators, constructs an aggregated service representation (e.g., service processing graph), performs necessary optimizations (e.g., function merging, redundant elimination), and calculates the best possible placement strategy by determining the PoPs to deploy the related VNFs and their chaining order.
The monitoring module is responsible for collecting statistics and events from both the service plane and the infrastructure, and for providing runtime feedback to other NFVO modules. Based on on-the-fly traffic conditions, the placement module can recalculate a new placement to improve performance. The scheduling module can dynamically make fine-grained scheduling decisions to attain resource efficiency. The scaling/failover module can also collaborate with the placement module to scale in/out particular VNFs or service chains to accommodate traffic fluctuations or to instantiate new VNF replicas upon failure. 
Based on the decisions made by the aforementioned modules, the NFVO closely interacts with the other two MANO plane components to realize the intended service configurations and resource allocations.

VIM is designed to configure infrastructure components to accommodate the heterogeneous VNFs or service chains instantiated in the service plane. In particular, it directs provisioning and release/upgrade of NFVI resources and manages the mapping between virtual and physical resources. It also manages the data path for network services by creating/deleting/updating virtual interfaces and logical links, and collects the NFVI software and hardware status on behalf of the NFVO monitoring module. Note that an instance of VIM might control all the resources of the whole NFVI or that of multiple NFVI-PoPs. In some cases, a VIM might control only a specific type of resource. 

On the other hand, VNFM interacts with the service plane and manages the instantiation, scaling, upgrade, and termination of individual VNFs and service chains. It also needs to synchronize with VIM to allocate or release the related infrastructural resources.
According to the ETSI specification, the MANO system may also maintain several data stores to store configuration information, such as network service descriptions, VNF templates, NFVI resources, etc.


\paragraph{From VM-centric MANO to Kubernetes-first lifecycle control.}
A notable evolution is that many CNF-oriented deployments aim to implement lifecycle management through Kubernetes-native control loops, rather than relying solely on specialized MANO workflow engines. In this view, the \emph{desired state} of a network service and its constituent CNFs is represented as Kubernetes resources (often via Custom Resource Definitions), while controllers/operators continuously reconcile the actual runtime state to match the desired intent. Classic MANO operations (e.g., instantiate, scale, upgrade, heal, and terminate) are thus realized through Kubernetes primitives (e.g., apply, rolling update, autoscaling) combined with operator logic. This line of work argues that such a refactoring is essential to become \emph{truly cloud native}, while still acknowledging that telecom workloads impose additional requirements not fully addressed by
generic cloud platforms. \cite{breitgand2021true_cloud_native_mano}


\subsection{NFV Infrastructure}
NFV Infrastructure (NFVI) comprises all the hardware and software components required to provision virtual network services. The infrastructure might belong to Internet service providers, cloud/edge operators, or simply infrastructure providers. It usually comprises a wide range of computing nodes and network equipment. Each computing node or network equipment is commonly referred to as NFVI-PoP. Network equipment in NFVI can be traditional purpose-built switches/routers or the emerging programmable switches that can be remotely orchestrated with SDN or P4~\cite{bosshart2014p4} semantics.
The most typical form of computing node in NFVI is the COTS servers. These servers typically contain several critical hardware components, including physical Network Interface Controllers (NICs), multicore CPUs, and main memory, and are interconnected via PCI buses.
The physical NICs are capable of operating at the Gigabit level, with multiple queues that promote parallelization. High-speed packet I/O techniques are also integrated by the NICs to transport packets to the service plane. Within the server, multicore CPUs are distributed across non-uniform memory access (NUMA) nodes to accelerate traffic processing. Aside from the CPU, other computing units, such as smartNICs and GPUs, are widely used in existing NFV platforms to further boost performance.
The virtualization layer in the COTS server provides the environment to accommodate network functions. Virtualization can occur at the hardware level, relying on bare-metal hypervisors, or at the OS level, using container engines. Some platforms even execute network functions as ordinary processes, which is addressed as Physical Network Functions (PNFs) in some works. In this paper, we refer to them universally as VNFs for simplicity.
To ensure efficient communication between the VNFs and the external network, virtual interconnects must be precisely configured. This is typically accomplished using state-of-the-art software, virtual switches, or customized forwarding tables. 
Note that we consider physical links between COTS servers and network equipment as part of NFVI as well.


\subsection{Service plane}
The service plane is populated with a variety of Virtual Network Functions (VNFs) that implement different processing to provide various network services. The distribution of VNFs inside virtual environments is quite flexible. For instance, a VNF or a whole service chain can be mapped to a single VM for execution, a VNF can also be split into finer-grained processing elements and deployed across multiple NFV PoPs. Also, VNFs are typically implemented using different programming abstractions and execute under different runtime execution models.


\section{Taxonomy of NFV platforms} \label{sec:taxonomy}
We devote this section to reviewing existing NFV platforms and classifying them according to their primary design purpose.
Based on our literature review and the reference architecture, we classify existing NFV platforms into three general design purposes: integrated NFV platform, MANO system, and NFVI optimization.
An integrated NFV platform comprises both the MANO plane and the NFVI to support either end-to-end service provisioning or VNF development.
A MANO system addresses all or a subset of management issues, such as instantiation/termination, placement, dynamic scaling, monitoring, and the resilience of network services.
An NFVI optimization platform strives to ensure an efficient packet data path by optimizing related procedures and eliminating redundant processing.


\subsection{Integrated NFV platform}
Integrated NFV platforms can be further classified into two categories. The first category is for network operators or service providers to efficiently provision end-to-end services for their clients or subscribers. The second category is intended to make VNF implementation less time-consuming and error-prone for network developers.

\subsubsection*{End-to-end service provisioning}
Several platforms were proposed for functionality implementation.
\textbf{UNIFY}~\cite{csaszar2013unifying} and \textbf{Cloud4NFV}~\cite{soares2015toward} are two earliest projects for integrated NFV platforms. UNIFY introduces a general framework to automate service provisioning. It employs a layered graph abstraction to automatically map user-specified services to the actual SFCs deployed in the underlying NFVI PoPs.
Cloud4NFV provides an SFC model to allow for fine-grained traffic classification and steering and relies on cloud management tools to perform NFV MANO operations.
\textbf{CloudBand}~\cite{cloudband} is an integrated NFV platform composed of a CloudBand node and a management system. The CloudBand node supplies resources to accommodate network services. The management system performs MANO operations across different service domains.
\textbf{GNF}~\cite{cziva2017container} brings NFV to the network edge. It exposes a graphical user interface for specifying service intent and displaying system events, and uses a manager to perform MANO operations. 
An agent is embedded into each edge device to manage the containerized VNFs. Given the resource constraints of edge devices, it runs VNFs in lightweight Linux containers rather than VMs.
{\bf DeepNFV}~\cite{li2018deepnfv} is built on top of GNF. It incorporates deep learning techniques inside VNFs to learn and discover hidden data patterns and provide enhanced services such as traffic classification, quality of service optimization, link status analysis, and so on. 
\textbf{NetFATE}~\cite{lombardo2015open} employs a similar architecture and deploys VNFs on both data center servers and edge devices, considering runtime traffic load and Quality-of-Experience (QoE).  
\textbf{SONATA}~\cite{draxler2017sonata} brings DevOps to NFV by providing a service development toolchain integrated with a service platform and orchestration system. The toolchain comprises a service-programming abstraction with supporting tools that enable developers to implement, monitor, and optimize VNFs or SFCs. The service platform includes a customizable MANO framework for deploying and managing network services. It also supports platform recursion and network slicing.
\textbf{Eden}~\cite{ballani2015enabling} is another platform proposed for provisioning network functions at end-hosts in a single administrative domain. It comprises a controller, stages, and enclaves on end-hosts. The controller provides centralized VNF coordination based on its global network view. Stages reside in the end-host stack to map application semantics to specific traffic classes. The per-host Eden enclave maintains a set of Match/Action tables to decide the destination VNF for each packet based on its traffic class. VNFs in Eden are written in F\#, compiled into executable bytecode, and then interpreted within the enclaves.

Other integrated NFV platforms focus on performance under high traffic loads (e.g., 40/100~Gbps).
\textbf{OpenBox}~\cite{bremler2016openbox} allows developers to implement VNF logic via the northbound API of the OpenBox controller, which then deploys the logic to the data plane and realizes the intended processing sequence via the OpenBox protocol. The OpenBox controller also merges the core control logic of multiple VNFs to avoid duplicate processing and free up NFVI-PoP resources for other tasks. The OpenBox data plane is extensible with specialized hardware or pure software. By default, OpenBox contains over 40 processing blocks that can be chained to realize various VNFs, and it can further seamlessly integrate custom blocks at runtime.
\textbf{Slick}~\cite{anwer2015programming} allows developers to write network functions in a high-level programming language and specify the intended network service for different traffic flows or classes. The Slick controller performs the placement and traffic steering using several heuristic algorithms. In particular, the Slick runtime parses the specified policy, determines the optimal servers on which to place the elements, and installs forwarding routes to the in-path switches to realize the intended processing sequence. 
A shim is configured on each server to provide up-to-date system status for the Slick runtime to make incremental optimizations. 
\textbf{Elastic~Edge~(E2)}~\cite{palkar2015e2} is meant for a general NFV platform that frees developers from common deployment and management issues, which are instead delegated to the E2 manager. E2 allows network operators to express network policies in terms of individual pipelets, each consisting of a subset of input traffic (or traffic class) and a processing graph. Similar to OpenBox, the E2 manager merges the pipelets into a single graph and instructs local agents to place the corresponding network functions across the server cluster and interconnect them through the high-speed E2 data plane. E2 also provides hooks for VNFs and the data plane to detect system overload and dynamically scale NF instances with flow affinity guaranteed.
\textbf{SDNFV}~\cite{zhang2016sdnfv} combines SDN and NFV to realize a flexible, hierarchical control framework over VNFs. It comprises three hierarchies: SDNFV application, SDN controller, NFV orchestrator, and NF manager. 
SDNFV application utilizes a graph abstraction to represent the intended network services for different traffic flows. Then it proposes a heuristic algorithm to jointly deploy VNFs on COTS servers and configure traffic routes across them via the SDN controller and NFV orchestrator. An instance of NF manager is installed on each COTS server to manage the local VNFs and traffic routing. Each manager maintains an extended OpenFlow (OF) table based on host-level status. This table can also be configured by the remote SDN controller (for default routing) and the local VNFs (based on their internal states), realizing a more flexible control paradigm beyond SDN.
\textbf{MicroNF}~\cite{meng2019micronf} addresses the consolidation, placement, scaling, and scheduling of modularized SFCs. It consists of a centralized controller and a high-speed infrastructure. The MicroNF controller uses a graph constructor to analyze inter-element dependencies and reconstruct the service graph to reuse redundant elements. 
Then it uses a placer to optimally place and consolidate modularized VNFs, reducing inter-VM data transfer. At runtime, it dynamically collects element statistics and applies two resource-scaling algorithms to SFC scaling with minimal inter-element latency. Also, the MicroNF infrastructure ensures high-speed, consistent packet forwarding and fair VNF scheduling. 
\textbf{$\mu$NF}~\cite{8806657} advocates designing network functions as Microservices, since the self-contained, loosely coupled design supports fine-grained resource allocation and VNF scaling. $\mu$NF is designed for building VNFs and SFCs using disaggregated, reusable network processing components. It consists of a centralized orchestrator and a cluster of per-server agents. Similar to other integrated NFV platforms, it allows network operators to specify service requests as directed graphs, which the orchestrator converts into equivalent forwarding graphs. Then, agents on the related COTS servers are instructed to deploy and interconnect the corresponding VNFs according to the graph specifications.
\textbf{Metron}~\cite{katsikas2018metron} leverages the resources of both the underlying hardware and the programmable network to achieve high-speed processing. The Metron controller parses the traffic classes associated with a service chain and generates a synthesized processing graph. Then it decomposes the graph into a set of stateless operations offloaded to in-path programmable network equipment, while mapping the remaining stateful operations to COTS servers. To reduce the overhead of software-based traffic dispatching, Metron leverages in-path programmable switches to tag packets, which the COTS NIC matches to dispatch to the correct cores for ``run-to-completion" processing. For management, Metron additionally deploys an agent on each server to monitor the runtime statistics and scale overloaded VNF instances accordingly.
\textbf{OPNFV}~\cite{price2012opnfv} is a joint open-source project to promote NFV deployment and innovation. It involves a large compilation of tasks, including continuous integration of components from upstream projects, function verification, performance benchmarking, and service automation. As an integrated platform, it provides full-fledged features, including VNF management, dynamic service provisioning, prompt fault detection and recovery, and vendor- and operator-agnostic deployment.
\textbf{NNF}~\cite{bonafiglia2016enabling} explores capabilities of resource-constrained devices to deploy VNFs. The authors present preliminary results demonstrating the feasibility of implementing NNF across a variety of servers with heterogeneous hardware specifications and accelerators.
\textbf{CoNFV}~\cite{xu2018confv} combines cloud and end-hosts to reduce service deployment cost and processing latency. An abstraction is proposed to divide SFC processing logics between cloud infrastructure and end-hosts. An intuitive API is also designed to enable developers to port existing VNFs to CoNFV.
Bento~\cite{reininger2021bento} brings NFV-style extensibility into Tor by enabling the deployment of
additional traffic-processing functions under strict safety constraints. It illustrates a domain-specific,
end-to-end NFV platform where the control objectives include not only performance and agility but also
strong security and correctness requirements for extensible processing.

\subsection{Cloud-deployable NFV platforms: Kubernetes-first CNFs}
\label{sec:taxonomy-cloud-native-nfv}

While early NFV platforms were predominantly VM-centric and often coupled with OpenStack-style infrastructure, a growing class of systems seeks to make NFV \emph{cloud-deployable} by leveraging cloud-native substrates wherever possible. The key idea is to treat the cluster manager (e.g., Kubernetes) and common cloud control mechanisms as the default building blocks, and then \emph{adapt} them to satisfy NFV-specific requirements such as service
chaining, strict performance isolation, SLO adherence, and rapid scaling.

Quadrant exemplifies this direction by explicitly positioning NFV for deployment in commodity clouds and by reusing common cloud infrastructure, such as Kubernetes and serverless-style mechanisms, while introducing NFV-specific components for packet processing, scheduling, isolation, and autoscaling. \cite{wang2022quadrant}

In parallel, the industry increasingly captures cloud-native NFV expectations through operator-facing principles and infrastructure reference frameworks. For example, the NGMN Cloud Native Manifesto summarizes key operator principles for cloud-native telecom systems, and the CNTT reference framework aims to standardize the underlying cloud infrastructure profiles used to host virtualized and containerized network workloads. These documents are not platform implementations themselves, but they shape the constraints and evaluation criteria that NFV platforms must satisfy in the CNF era. \cite{ngmn_cloud_native_manifesto,cntt_whitepaper}


\subsubsection*{VNF development}
There is a collection of works dedicated to facilitating the development of VNFs. They aim to provide high-level APIs to facilitate VNF development. Most of them are equipped with runtimes to guarantee efficient VNF execution. The spirit of these works is to free developers from reinventing the wheel for common management tasks and to let them focus on implementing VNF control logic.
\textbf{xOMB}~\cite{anderson2012xomb} is the earliest effort to build scalable, programmable, and high-performance middleboxes on COTS servers. It arranges a set of programmable modules into a general pipeline to implement the expected network function. The xOMB control plane monitors the execution pipeline to make timely adjustments when instances fail.
\textbf{CoMb}~\cite{sekar2012design} advocates consolidating network functions at both the execution and management levels. The centralized CoMb controller takes as input the service policies and infrastructure specifications and solves an optimization model to decide the optimal deployment strategy, which is mapped to the distributed CoMb middleboxes by allocating the required resources.
\textbf{FlowOS}~\cite{bezahaf2013flowos} is a kernel-based programmable platform purposed for middlebox development. It implements an API to facilitate flow processing in VNFs and to conceal low-level complexities such as Inter-process communication, low-level packet delivery, and synchronization.
\textbf{NetBricks}~\cite{panda2016netbricks} facilitates VNF development by providing a small set of highly optimized, customizable core processing elements implemented as user-defined functions. Instead of relying on VMs or containers, NetBricks employs safe language, an efficient runtime library, and unique types to ensure a similar level of VNF isolation with much lower overhead.
\textbf{Scylla}~\cite{riggio2016scylla} is a declarative language for flow-level VNF development. It provides several high-level abstractions to allow developers to express their intents, such as SFCs, VNF monitoring, and state management. The Scylla runtime is responsible for fulfilling these intents within the infrastructure.
\textbf{libVNF}~\cite{naik2018libvnf} implements a generic library to assist the development of VNFs ranging from L2/L3 middleboxes to transport-/application-layer endpoints, with the support of seamless integration of the kernel and third-party network stacks. A request object abstraction is proposed to maintain application states across multiple non-blocking, event-driven callbacks. The libVNF API also supports interacting with multi-level data stores for state management across threads within a single VNF or across multiple VNF replicas.
\textbf{Flick}~\cite{alim2016flick} brings application-specific semantics into VNF development on multi-core COTS servers. The authors implement a domain-specific language named \emph{flick} that offers high-level abstractions and common primitives to assist VNF development. The compiler automatically translates the flick programs into parallel task graphs with bounded runtime resource usage. Multiple graphs can execute simultaneously without interference through cooperative scheduling.
\textbf{NFMorph}~\cite{alipourfard2018decoupling} proposes to decouple network function logic from packet processing optimization. VNF programs are expressed in a domain-specific language with primitives that are coherent with packet processing pipelines. NFMorph runtime constructs a near-optimal processing pipeline incrementally from a preliminary pipeline using code profiling, traffic sampling, and system constraints. 
\textbf{Polycube}~\cite{miano2019service} is proposed to build VNFs that can be dynamically adjusted at runtime. Its kernel fast path leverages extended Berkeley Packet Filter (eBPF)~\cite{ebpf} to sustain high-speed packet processing. It also exposes a set of abstractions to ease SFC development.
The following platforms adopt different means to avoid the complexities introduced by the callback-based asynchronous programming model.
\textbf{NetStar}~\cite{duan2019netstar} implements a flow-based asynchronous interface combined with the future/promise C++ library for VNF development. Instead of spreading control logic across multiple callback functions, NetStar mimics sequential execution by chaining multiple future objects and continuation functions over a single function call. 
\textbf{ClickNF}~\cite{gallo2018clicknf} augments the Click modular router~\cite{kohler2000click} with a modular, configurable, and extensible TCP stack to build L2-L7 network functions. It further incorporates DPDK for kernel-bypassing packet I/O. A blocking I/O primitive is also proposed to alleviate development difficulty imposed by the traditional asynchronous non-blocking I/O paradigm.
\textbf{Rubik}~\cite{li2021rubik} elevates NFV programmability by treating a middlebox stack as a composable program rather than a hand-wired datapath. It provides abstractions for constructing and
transforming middlebox pipelines while keeping the resulting service easier to reason about (e.g.,
composition correctness and systematic rewrites).
\textbf{S6}~\cite{woo2018elastic} extends the distributed state object (DSO) with a programming model to build elastically scalable VNFs. The S6 runtime manages the shared VNF states distributed in the DSO space. To meet performance requirements, S6 employs a set of optimizations, including micro-threaded scheduling and DSO space reorganization.
\textbf{StatelessNF}~\cite{kablan2017stateless} embraces the separation of concerns design by decoupling the VNF states from processing so that developers only need to concentrate on VNF-specific logic, while StatelessNF arranges for state replication and management tasks. The VNF states are maintained in a distributed key-value data store that guarantees low-latency access and data resilience. The orchestration plane dynamically monitors the status of VNFs and infrastructure and makes adjustments.
\textbf{NFVactor}~\cite{duan2019nfvactor} tackles efficient flow migration and failure recovery. It employs a distributed actor model with low processing overhead and provides APIs for building VNFs with resilience guarantees. The per-flow actors can perform flow migration and replication in parallel while NFVactor runtime timely schedules each actor and guarantees high-performance execution.


\subsection{Management and Orchestration}
Many platforms are purpose-built to provide NFV MANO solutions. Some of them strive for full-fledged, holistic MANO systems, while others tackle only a subset of MANO issues, such as scheduling, monitoring, scaling, load balancing, failover, and VNF and SFC management. In this section, we discuss these two categories of solutions respectively.

\subsubsection*{Holistic MANO system}
\textbf{ETSO}~\cite{mechtri2017nfv} is an ETSI-compliant NFV MANO platform for end-to-end SFC provisioning over heterogeneous cloud environments. It addresses key NFV orchestration challenges with a shared service abstraction and optimal VNF/SFC placement algorithms across heterogeneous NFV, SDN, and cloud computing technologies.
\textbf{OpenMANO}~\cite{lopez2015openmano} aims to implement the ETSI NFV MANO framework with guaranteed performance and portability. It consists of a MANO system (openmano), a virtual infrastructure manager (openvim), and a graphical user interface (GUI). The openvim directly manages the NFVI resources. It also interacts with an SDN controller to establish the intended traffic data path and relies on a REST northbound API to communicate with the openmano, where relevant MANO tasks are performed.
\textbf{Open Baton}~\cite{openbaton} is another ETSI MANO-compliant platform with the major objective of developing an extensible and customizable framework for service orchestration across heterogeneous NFV Infrastructures. It manages a diverse set of VNFs running on a multi-site NFVI with heterogeneous virtualization technologies. It also features network slicing using SDN technologies to multiplex the infrastructural resources across multiple VNF instances or network services.
\textbf{vConductor}~\cite{shen2015vconductor} supports completely automated virtual network deployment by simplifying the
provisioning procedure. It also adopts a multi-objective resource scheduling algorithm to meet individual business requirements and employs enhanced inventory management to facilitate fault isolation. It further employs a plug-in architecture with a modular design to enhance interoperability with various NFV management entities.
\textbf{T-NOVA}~\cite{xilouris2015t} leverages SDN controllers and cloud management tools to design and implement a software NFV MANO stack to automate the deployment and management of Network Functions-as-a-Service (NFaaS) over virtualized network infrastructures. In \cite{trajkovska2017sdn}, an open-source toolkit is implemented and evaluated using T-NOVA for end-to-end service provisioning in Datacenter-based NFVI.
{\bf TeNOR}~\cite{riera2016tenor} is an NFV orchestrator based on microservice architecture. It proposes two approaches to address resource and service mapping.
{\bf 5TONIC}~\cite{nogales2019design} is an open-source NFV MANO platform supporting secure cross-site message exchanges for control plane and data plane. It can provide experimental environments to multiple authorized parties with pre-defined security policies.


\subsubsection*{Scaling and failover}
One attractive aspect of NFV is its flexibility in handling traffic and system dynamics. When input traffic swells, the MANO plane can dynamically launch new instances to rebalance the workload. When a VNF instance crashes, new replicas can also be promptly initiated to avoid service disruption.
Correspondingly, several platforms endeavor to guarantee efficient scaling and failover, especially for stateful VNFs.
\textbf{Split/Merge}~\cite{rajagopalan2013split} exposes a programming abstraction that promises transparent, load-balanced VNF scaling with guaranteed per-flow state consistency. In its prototype system, a centralized orchestrator, along with an SDN controller, is employed to direct VNF scaling and flow migration. A VMM agent is deployed on each server to create or remove VNFs on demand. By integrating the Split/Merge API, per-flow VNF states are Split or Merged across multiple replicas. The system then migrates the relevant states and configures the network to direct flows to the correct replicas.
\textbf{TFM}~\cite{wang2016transparent} also aims at achieving safe, transparent, and efficient flow migration. Similar to Split/Merge, it instructs the migration process through a centralized controller. The controller decouples flow and state migration processes with three modules: a state manager, a flow manager, and a forwarding manager. The state manager conducts state migration through southbound APIs. The forwarding manager interacts with the SDN controller to set traffic steering rules. The flow manager manages the TFM boxes, which perform packet classification and buffering during flow migration.
\textbf{OpenNF}~\cite{gember2014opennf} is a non-intrusive control framework that guarantees SLAs, ensures correct packet processing, and uses resources efficiently. It implements a controller comprising an event-driven model to capture relevant packets, a southbound API to request the import/export of VNF states at different granularities (i.e., single/multiple/all flows), and a northbound API for control applications to instruct state migration or synchronization. In particular, state migration is carefully crafted to avoid packet losses or out-of-order processing, while state synchronization can be performed with either strong or eventual consistency.
\textbf{DiST}~\cite{kothandaraman2015centrally} and \textbf{U-HAUL}~\cite{u-haul} follow similar procedures for state and flow migration, but do not require the involvement of the centralized controller. 
For instance, U-HAUL aims at implementing an efficient state migration framework. Based on the observation that mouse flows are usually short-lived and that state/flow migrations incur high overhead, it identifies and migrates only the execution states of elephant flows, while keeping the states of mouse flows in the original VNF instance until expiration. Reducing the number of migrated flow states makes the migration process more efficient.
{\bf LEGO}~\cite{zhang2019nfv} is designed to scale VNFs empowered by Artificial Neural Network (ANN). It proposes a traffic-splitting scheme that splits incoming packets with ideal granularity. It also partitions ANN-based VNFs into smaller blocks to smooth the VNF scaling process. The centralized LEGO controller continuously monitors resource utilization across blocks to achieve resource efficiency by dynamically replicating or merging them.
{\bf Lange et al.}~\cite{lange2019machine} proposes a machine learning approach to adapt the number of VNF replicas based on recent monitoring data. The authors also propose a workflow to generate labeled training data that can reflect the real-world network dynamics.
\textbf{StateAlyzr}~\cite{khalid2016paving} is a non-intrusive framework that automatically handles state clone and migration based on program analysis. It employs program analysis techniques and three algorithms to automatically identify all relevant VNF states that need to be redistributed, reducing the target codebase and porting effort. Specifically, StateAlyzr undertakes a three-stage procedure: it begins by identifying the per-/cross-flow state, based on which it filters the updatable ones and identifies the relevant header fields. 
\textbf{CHC}~\cite{khalid2019correctness} adopts a set of state management and optimization techniques to ensure service correctness without degrading performance. In particular, it offloads VNF states to the distributed data store and employs state caching and update algorithms to ensure high performance. It additionally leverages metadata to ensure a set of correctness properties during traffic redistribution and in the presence of instance/component failures.
\textbf{SFC-Checker}~\cite{tschaen2016sfc} is a diagnosis framework to verify the correctness of SFC forwarding behaviors. It extends OpenFlow to represent each VNF with a Match/Action table and a state machine, and builds a stateful forwarding graph to capture both forwarding behaviors and state transitions, enabling verification of SFC forwarding behaviors under different traffic conditions.
\textbf{FTMB}~\cite{sherry2015rollback} adopts a log and rollback recovery scheme to conduct VNF failover using log replay. It incorporates two approaches, ordered logging and parallel release, to retain correctness and high-performance rollback.


\subsubsection*{Scheduling}
Some NFV platforms are specialized for VNF scheduling to achieve performance isolation or resource efficiency.
\textbf{NFVnice}~\cite{kulkarni2017nfvnice} is an NFV framework providing fair scheduling and efficient chaining. In particular, it adopts rate-cost proportional fairness by adjusting each VNF's CPU weight based on its estimated arrival rate and service time. The scheduling is performed by tuning the OS scheduler via Linux cgroups. In addition, it actively monitors VNF backlogs and employs a back-pressure mechanism to early-drop packets early for each congested service chain to spare resources. It also comprises an asynchronous scheme to multiplex I/O with processing.
\textbf{EdgeMiner}~\cite{zhang2019characterizing} seeks to reuse the spare CPU resources of VNFs to execute data-processing applications at the network edge. EdgeMiner uses interrupt-based I/O for VNFs instead of polling to save CPU resources under low workload. It also employs a back-pressure scheme to dynamically detect service chain overloads and puts upstream VNFs into sleep to harvest the otherwise wasted CPU cycles.
\textbf{UN$i$S}~\cite{chowdhury2018unis} is a scheduling system tailored for poll-mode DPDK-based VNFs. For each worker core, it subsequently retrieves statistics from the inter-VNF buffer of the assigned SFCs and makes scheduling decisions based on buffer occupancy. The scheduling is non-intrusive, as UN$i$S tunes only the parameters of the Linux Realtime Scheduling without rewriting the VNFs.
\textbf{SNF}\footnote{The SNF here is not the same as the other SNF~\cite{katsikas2016snf}}~\cite{singhvi2019snf} leverages serverless computing for stateful VNFs. It dynamically traces the workload demands of each VNF and allocates compute resources with fine-grained granularity. A peer-to-peer in-memory store is deployed to proactively replicate states and reduce packet-processing latency.
\textbf{ResQ}~\cite{tootoonchian2018resq} is a cluster-based resource management framework that achieves guaranteed service-layer objectives. It consists of a performance profiler and a scheduler. The profiler performs a set of experiments on the target VNFs to construct profiles. Based on the profiling results, the ResQ scheduler computes a resource-efficient allocation using a greedy approach. ResQ also periodically solves a Mixed-Integer Linear Programming (MILP) formulation to obtain the optimal allocation, which can replace the current allocation if a pre-defined threshold is exceeded.
\textbf{NetContainer}~\cite{hu2017towards} aims at exploiting cache locality to achieve maximum throughput and low latency for containerized VNFs. The authors first identify the random page allocation policy as the root cause of cache pollution. Then they build an estimation model based on the footprint theory to infer the cache access overhead and model the cache mapping problem as a Minimum Cost/Maximum Flow (MCMF) problem to decide the optimal memory buffer mappings.
\textbf{NFV-throttle}~\cite{cotroneo2017nfv} controls VNF overload by spreading software modules in NFV infrastructure. These modules dynamically monitor system conditions and selectively drop excessive packets to prevent VNFs from being overwhelmed.
To ensure strict processing isolation for co-located containerized VNFs, \textbf{Iron}~\cite{khalid2018iron} introduces an enforcement mechanism to account for the time each VNF spent in the kernel stack. Then it throttles, or even drops, packets for aggressive VNFs using either a Linux scheduler or a hardware-based approach.


\subsubsection*{Profiling}
NFV complicates monitoring and troubleshooting because packet processing is distributed across multiple VNFs that may share CPU cores, memory hierarchy, and I/O resources. As a result, end-to-end performance degradation can stem from (i) contention among co-resident VNFs, (ii) queueing and backpressure across a service chain, or (iii) insufficient visibility into per-flow dynamics at high speed.
Recent systems, therefore, treat observability and diagnosis as first-class enablers for MANO decisions (e.g., consolidation, placement, and scaling).

Several platforms are dedicated to performance characterization.
\textbf{NFV-vital}~\cite{cao2015nfv} is among the earliest efforts towards VNF performance characterization.
It implements four components that can be seamlessly integrated into an ETSI-compatible NFV platform. With NFV-vital, users can specify their deployment and workload configurations, which the NFV-vital orchestrator interprets to set up VNFs and generate the workload accordingly. NFV-vital orchestrator receives statistics at runtime and performs post-test analysis. A similar design pattern is also adopted by several other platforms.
{\bf Gym}~\cite{rosa2017take} is designed for automatic VNF performance benchmarking. It embraces a modular architecture with an extensible set of benchmarking tools and a simple messaging subsystem for remote procedure calls. It further provides a means for data post-processing and visualization of results.
{\bf Du et al.}~\cite{du2018service} builds a benchmarking framework on OPNFV clearwater platform. They leverage a microservice architecture to integrate existing open-source tools, enabling comprehensive testing under varied traffic loads and fault conditions.
\textbf{ConMon}~\cite{moradi2017conmon} is a distributed framework to monitor the performance of containerized VNFs. It dynamically discovers and monitors communication between containers and executes network monitoring functions within a standby container interconnected via a virtual switch.
\textbf{Symperf}~\cite{rath2017symperf} predicts VNF runtime performance and functional behaviors under various traffic dynamics through code analysis.
\textbf{KOMon}~\cite{geissler2019komon} is a kernel-based online monitoring tool to measure packet processing times imposed by the target VNF.
Rather than focusing solely on individual VNFs, several platforms can characterize SFC performance. For example, \textbf{SFCPerf}~\cite{sanz2018sfcperf} is proposed for automated SFC performance evaluation. Similar to NFV-vital, it has a control module that receives a user-specified configuration file and deploys the corresponding service chain in the target infrastructure. The control module subsequently collects data for analysis and visualization.
\textbf{NFVPerf}~\cite{naik2016nfvperf} is able to detect performance bottlenecks on a service chain by monitoring inter-component communication. \textbf{Perfsight}~\cite{wu2015perfsight} also aggregates execution information from various components along the data path to diagnose performance issues. \textbf{VBaaS}~\cite{rosa2015vbaas} profiles SFC performance for distributed NFVI. certification of VNFs and NFVI PoPs as performance profiles; fine-tuning choices of VBaaS results with tailor-made PoPs. VBaaS processes can be defined as simple tasks in common workflows of collection, synchronization, integration, and export for benchmark analysis. Locations where these tasks can take place would be defined according to VBaaS profiles, orchestrator policies, and infrastructure capabilities

However, none of the aforementioned work is suitable for measuring and analyzing the performance of high-speed VNFs operating at multi-gigabyte rates. 
{\bf FloWatcher-DPDK}~\cite{zhang2018high,zhang2018flowmon,zhang2019flowatcher} is a lightweight DPDK-based flow monitor that targets
\emph{line-rate, flow-level} visibility in software. Instead of relying on coarse host counters, it focuses on collecting fine-grained per-flow statistics with low overhead, making it a practical building block for profiling and validating DPDK-accelerated VNFs and traffic generators under high-speed workloads.
\textbf{OPNFV Barometer}~\cite{barometer} is designed to monitor the performance of DPDK-accelerated VNFs. It can be attached to the target VNF as a secondary process to gather shared processing information. 
\textbf{NFV-VIPP}~\cite{dodarenfv} can be integrated into the DPDK-accelerated data plane to collect execution metrics and demonstrate the internals of an NFVI node.
\textbf{BOLT}~\cite{iyer2019performance} defines the performance contract, which expresses the expected VNF or SFC performance as a function of critical parameters (e.g., execution instructions, CPU cycles, memory accesses).
\textbf{DeepDiag}~\cite{gong2019deepdiag} monitors the runtime queuing statistics for each VNF and constructs an online impact graph to diagnose the cause of performance degradation.
\textbf{CASTAN}~\cite{pedrosa2018automated} can parse VNF code and automatically generate worst-case workloads that degrade performance. It adopts symbolic execution to identify the worst code path and a CPU cache model to determine the specific memory access pattern that causes L3 cache invalidation. According to the paper, CASTAN has successfully analyzed a dozen DPDK-based network functions.
{]bf Microscope}~\cite{microscope_sigcomm20} focuses on diagnosing performance problems \emph{inside} a chain of network functions.
Rather than treating VNFs as opaque boxes, it leverages queueing signals to localize bottlenecks and attribute end-to-end degradation to specific stages or resources, turning troubleshooting into a structured workflow.
{\bf AuditBox}~\cite{liu2021auditbox} adds an accountability facet to NFV chaining: it produces verifiable evidence about how traffic traverses a service chain, enabling after-the-fact validation of chain execution. This shifts “monitoring” from purely performance/health telemetry toward evidence of correctness, which is especially relevant for multi-tenant and multi-stakeholder NFV settings.
{\bf PIX}~\cite{iyer2022pix} represents a troubleshooting-oriented MANO facet: it argues that platforms need explicit performance interfaces that expose actionable internal signals for locating where time/resources are spent across NFs and chains. Rather than relying purely on external throughput/latency observations, PIX makes diagnosis NF-aware and chain-aware, enabling systematic bottleneck localization in consolidated deployments.
{\bf Klint}~\cite{pirelli2022klint} is a verification-oriented facet that complements runtime monitoring: it automates the checking of network function binaries against intended properties, reducing the operational risk of deploying opaque or third-party NF implementations.
To mitigate the impact of data collection on the data plane, a line of work explores an infrastructural-level feature to derive performance insights. In particular, 
{\bf SLOMO}~\cite{slomo_sigcomm20} targets a common root cause of NFV performance unpredictability: co-located VNFs contend for shared hardware resources (in particular, multiple components of the memory subsystem), resulting in non-obvious throughput/latency degradation. It models each NF's \emph{contentiousness} and \emph{sensitivity} and predicts contention-induced slowdown, enabling orchestrators to make better consolidation and provisioning decisions under SLO/SLA constraints.
{\bf Shelbourne et al.}~\cite{shelbourne2019learnability, shelbourne2021inference} approach the performance monitoring of high-speed NFs by inspecting.


\subsubsection*{Secure execution}
There is another group of platforms specifically devoted to developing secure VNFs for execution in untrusted environments.
\textbf{vEPC-sec}~\cite{raza2019vepc} incorporates a variety of traffic encryption, validation, and monitoring schemes to safeguard cloud-based LTE VNFs.
\textbf{SplitBox}~\cite{asghar2016splitbox} distributes VNF functionalities to multiple cloud VMs to obscure its internals from the public cloud.
\textbf{Embark}~\cite{lan2016embark} allows VNFs to operate on encrypted data, leveraging a special HTTPS encryption scheme.
\textbf{BSec-NFVO}~\cite{rebello2019bsec} introduces a blockchain-based architecture to protect NFV orchestration by auditing all the operations over the SFCs.
Other platforms exploit Intel\textsuperscript{\textregistered} Software Guard Extensions (SGX)~\cite{sgx} instruction codes to secure VNFs from memory reading attacks. 
In specific, \textbf{S-NFV}~\cite{shih2016s} concentrates on the protection of VNF states by stashing them into the shielded SGX memory region (\emph{enclave}) to prevent unauthorized access or snooping.
\textbf{TrustedClick}~\cite{coughlin2017trusted} and \textbf{ShieldBox}~\cite{trach2018shieldbox} extend the Click modular router to secure packet processing within SGX enclave, and rely on SGX remote attestation to verify code correctness. ShieldBox additionally integrates DPDK for high-speed packet processing and ring buffers to support SFC deployment. However, neither of them protects VNF states.
Building on these prior endeavors, \textbf{SafeLib}~\cite{marku2019towards} aims to provide a generic platform that offers comprehensive protection for VNFs, including user traffic, VNF code, policies, and execution states. The authors propose integrating DPDK and libVNF to support TCP functionality without compromising performance. Its implementation is currently underway.
\textbf{Safebricks}~\cite{poddar2018safebricks} and \textbf{LightBox}~\cite{duan2019lightbox} also strive for comprehensive protection while sustaining reasonable performance.
LightBox implements a virtual interface to ensure secure packet VNF I/O in SGX enclaves and adopts the mOS stack to support stateful VNFs. It also implements a flow-state management scheme by caching the states of active flows within enclaves while encrypting and storing the states of other flows on the untrusted host. A space-efficient hash algorithm is also incorporated for efficient flow classification.
SafeBricks relies on the NetBricks platform for packet processing and VNF code protection. It partitions VNF code to minimize the trusted computing base within enclaves and performs packet exchanges across the trust boundary via a shared-memory mechanism. It also supports deploying an entire SFC inside an enclave and leverages Rust primitives to isolate the VNFs.
Note that it is unclear if SafeBricks supports TCP functionalities.


\subsection{NFVI acceleration}
With the rapid growth in traffic volume, NFV platforms are expected to deliver services without compromising performance, even at high data rates. As the data plane, NFVI is inarguably the principal point of optimization. Many endeavors are concentrating on NFVI acceleration.


\subsubsection*{Single NFVI-PoP optimization}

\textbf{NetVM}~\cite{hwang2015netvm} aims to achieve high performance, flexible deployment, easy management, and guaranteed security. It achieves line-rate processing through a zero-copy packet delivery mechanism based on shared memory and relies on a hypervisor-based switch to flexibly steer traffic between VNFs and the network. To ensure security, NetVM defines multi-level trust domains to limit the memory access of untrusted VNFs. A control plane is also implemented to facilitate system management by making decisions either locally or remotely.
\textbf{OpenNetVM}~\cite{zhang2016opennetvm} is based on the NetVM architecture but uses lightweight Docker containers to wrap VNFs. It also enables more flexible traffic steering, as both the VNFs and the management entities can make routing decisions.
{\bf NetML}~\cite{dhakal2019netml} is built based on OpenNetVM and runs machine learning applications as VNFs. To offload computation to the GPU and unburden the CPU, NetML further extends the CUDA library to accelerate traffic processing.
\textbf{ClickOS}~\cite{martins2014clickos} aims to provide a high-performance, flexible, and scalable NFV platform with guaranteed resource/performance isolation and multi-tenancy. It utilizes the Click Modular Router~\cite{kohler2000click} to build a wide range of VNFs in Xen-based uni-kernel VMs. A set of optimizations is applied to the hypervisor data path to improve performance. The ClickOS VMs are instantiated from small images and boot in milliseconds.
Similarly, \textbf{HyperNF}~\cite{yasukata2017hypernf} aims at achieving high performance and resource utilization. It advocates consolidating VNFs to share CPU cores and uses hypervisor-based virtual I/O to reduce synchronization overhead.
Another platform with a similar design architecture is \textbf{CliMBOS}~\cite{gallo2018climbos}, which is based on Xen and ClickNF~\cite{gallo2018clicknf}. It is devised to help developers construct lightweight, isolated, and modular IoT backends.
\textbf{MVMP}~\cite{zheng2018flexible} is also based on DPDK and containers. It uses a virtual device abstraction layer to multiplex physical NICs and steer traffic.
\textbf{NFF-Go}~\cite{nff-go} is designed to build and deploy network functions in the cloud. It leverages DPDK to accelerate packet I/O and uses the Go language to facilitate development. The Go language is also expected to improve concurrency and guarantee safety. Moreover, it includes a scheduler that dynamically scales packet processing based on the current workload.
{\bf LemonNFV}~\cite{li2023lemonnfv} targets consolidation as an NFVI optimization problem. It improves how
heterogeneous NFs share server resources so that packing does not collapse performance for demanding
workloads. 
{\bf Rashelbach et al.}~\cite{rashelbach2022scaling} focus on the virtual-switch datapath as the principal optimization target. It accelerates common-case packet processing by reusing previously computed results along the OVS processing path.

Some platforms mainly focus on SFC deployment and optimization.
In \textbf{Flurries}~\cite{zhang2016flurries}, the authors introduce the concept of per-flow service provisioning and implement a container-based NFV platform named Flurries that provides flexible service function chaining. In Flurries, each flow can be allocated to a corresponding SFC to allow for flow-level service customization and isolation. A combination of polling (for NICs) and interrupt (for VNFs) I/O scheme is also implemented to consolidate a large number of per-flow service chains on the server.
\textbf{Microboxes}~\cite{liu2018microboxes} concentrates on transport- and application-layer protocol consolidation for SFCs.
It implements a modular, asynchronous TCP stack that can be customized on a per-flow basis to avoid redundant protocol processing on a service chain. It also provides a publish/subscribe-based communication mechanism for chaining network functions and realizing complex network services.
\textbf{SNF}~\cite{katsikas2016snf} is also implemented to eliminate redundant processing for SFC. It uses graph composition and set theory to determine the traffic classes of incoming packets before synthesizing per-class, functionally equivalent, and optimized network functions.
Likewise, \textbf{NFCompass}~\cite{hu2018enabling} strives to shorten the service chain by synthesizing VNFs and exploring parallel execution. It additionally implements a graph-based model to minimize data transfer and balance traffic load.
\textbf{SpeedyBox}~\cite{jiang2019speedybox} utilizes a table-based Match/Action technique to consolidate VNF actions at runtime and eliminate redundant processing along a service chain.
\textbf{ParaBox}~\cite{zhang2017parabox} explores to shorten SFCs by exploring parallelized VNF execution. It consists of a dependency analysis module to determine if some VNFs can run in parallel, mirror/merge functions to distribute and aggregate packet copies across the parallelized VNFs.
Similarly, \textbf{NFP}~\cite{sun2017nfp} also aims at exploring VNF parallelism for SFCs. It allows network operators to express SFC intent as policies and implements an orchestrator to analyze VNF dependencies and compile the policies into optimized network service graphs. The NFP infrastructure sequentially executes the service graphs and handles low-level details such as traffic steering, load balancing, and parallel VNF execution.
\textbf{Dysco}~\cite{zave2017dynamic} explores a session protocol mechanism for runtime reconfiguration of TCP service function chains with no packet loss and minimal service interruption. 
\textbf{MiddleClick}~\cite{barbette2018building} aims at building high-speed, parallelized service chains. It allows network operators to define SFC intents, which are synthesized into a flow table and managed by the framework. A session abstraction is also implemented to facilitate per-flow inspection.
\textbf{ESFC}~\cite{shen2018nfv} is designed for flexible SFC resource management. It implements a controller to monitor the VNF status and enforce resource allocation policies using an asynchronous notification mechanism. A hash algorithm is devised to balance packets across VNF replicas while ensuring flow-level affinity.
\textbf{SCC}~\cite{katsikas2017profiling} collects system statistics such as hardware/software performance counters, to identify the root causes of excessive SFC delays. Based on the profiling results, the SCC runtime addresses performance bottlenecks by applying a set of optimizations, including tuning the I/O batch size and adjusting scheduling policies, priorities, or time slices for the SFCs.
{\bf PA-Flow}~\cite{taguchi2019fast} undertakes a packet aggregation approach to reduce traffic load for each VNF on an SFC. It embeds a module on each NFVI-PoP to perform transparent, network-aware, hop-by-hop packet aggregation/disaggregation to realize high-speed SFCs. Note that PA-Flow is designed to be compatible with state-of-the-art virtualization techniques and packet I/O frameworks.
{\bf Zorello et al.}~\cite{zorello2018improving} develop an in-path prediction engine to save energy for cloud-based NFVI. The proposed method employs a pre-trained machine learning model with Dynamic Voltage-Frequency Scaling (DVFS) to predict and specify the most suitable CPU frequency at runtime. 
{\bf HALO}~\cite{yuan2019halo} is designed to perform near-cache flow classification in software virtual switches with minimal impact on collocated VNFs. It explores last-level cache parallelization through the Non-Uniform Cache Access (NUCA) and Caching/Home Agent (CHA) on Intel multicore CPUs. The authors also extend the OS instruction with three table lookups to utilize the proposed HALO accelerators associated with each CHA.
{\bf Maestro}~\cite{pereira2024maestro} targets multicore scaling of software network functions as a practical NFVI-side acceleration problem: it reduces the manual effort of restructuring sequential NF implementations to
exploit parallel execution while preserving packet/flow semantics. It fits the single-PoP optimization branch
because it directly raises throughput on commodity servers without changing the service’s external behavior.
{\bf NFOS}~\cite{yan2024transparent} addresses a complementary scaling scenario: it aims to accelerate real-world, often single-threaded NFs by transparently exploiting multicore servers while preserving expected behavior. It raises achievable throughput without requiring invasive rewrites of deployed NF logic.
{\bf Bansal et al.}~\cite{bansal2023disaggregating} revisits state as the central obstacle to elastic scaling: it separates packet-processing compute from state management, allowing compute resources to be provisioned and scaled more flexibly while state is handled by a dedicated substrate. 
{\bf Sirius}~\cite{gao2024sirius} represents hybrid acceleration, where parts of a service chain are mapped into
P4-capable gateways and the remainder stay in software. Rather than only accelerating a single NF, it
treats the chain as the unit of mapping across heterogeneous dataplanes.
SyNAPSE~\cite{pereira2022automatic} reduces the engineering barrier of NFV acceleration by automatically
generating network-function accelerators from component-level building blocks. It shifts acceleration from manual one-off implementations toward repeatable,
toolchain-driven generation across targets.


\subsubsection*{Hardware-assist design}
\textbf{P4SC}~\cite{chen2019p4sc} and \textbf{P4NFV}~\cite{he2018p4nfv} explore P4 language to accelerate SFC processing.
P4SC leverages the P4 to construct and consolidate SFCs. It parses SFC policies specified by network operators and converts them into a P4 program, which is subsequently deployed on P4-compatible hardware. P4NFV is designed for both hardware and software targets and supports runtime reconfiguration without violating state consistency.
Albeit augmented with various software acceleration techniques, CPU cores might still fall short of performance. As a result, several platforms explore other hardware components for processing acceleration.
\textbf{OpenANFV}~\cite{ge2014openanfv} features automated provisioning and elastic management of network services. To address performance issues caused by computation-intensive VNFs, it offloads a subset of functionality to programmable hardware.
\textbf{UNO}~\cite{le2017uno} targets SmartNICs (i.e., ASIC, FPGA, System-on-Chip) for computation offloading without violating interoperability with the existing orchestration plane. While still relying on a centralized orchestrator to make global decisions, UNO selectively places new VNFs on the underlying SmartNICs to minimize host CPU usage, using a placement algorithm that considers local system status. It also actively reruns the algorithm and adjusts VNF placement between the host and SmartNICs. To hide the complexity of SmartNICs from the remote orchestrator, UNO exposes a single-switch abstraction that correctly maps, ensuring rules to the host or SmartNIC switches.
\textbf{NICA}~\cite{eran2019nica} is a hardware-software co-designed platform for inline data path acceleration on SmartNICs with integrated FPGAs (F-NICs). The platform exposes a programming abstraction that gives applications direct control over F-NIC accelerators and an I/O path virtualization that allows multiple VMs to share the F-NIC with guaranteed security and fairness.
{\bf Nezha}~\cite{li2025nezha} advances SmartNIC-assisted NFVI acceleration by treating on-NIC resources
as a poolable substrate for virtual-switch processing. It studies how SmartNIC constraints can become the
bottleneck and proposes pooling/sharing strategies to improve performance isolation and throughput for
demanding tenants.

Another line fo works, including \textbf{FlowShader}~\cite{yi2019flowshader}, \textbf{GPUNFV}~\cite{yi2017gpunfv}, \textbf{Gen}~\cite{zheng2018gen}, \textbf{Grus}~\cite{zheng2018grus}, and \textbf{G-NET}~\cite{zhang2018g}, seeks to construct a concerted CPU-GPU pipeline to expedite SFC processing.
FlowShader explores GPU and CPU for paralleled processing. It leverages the standard Linux TCP/IP stack to classify incoming traffic, and specialized data structures to buffer messages and maintain per-flow information. After a batch completes, it invokes a flow-scheduling algorithm to balance buffered data between the GPU and CPU for processing. Notably, each GPU thread executes the entire processing logic of a VNF (or SFC) to ensure flow-level parallelism. FlowShader further provides a general API for developing compatible VNFs across the CPU and GPU domains.
GPUNFV also employs flow-level parallelism and wraps an entire SFC within a single GPU thread. Compared with FlowShader, it only exploits the GPU for processing, while devoting the CPU to kernel-bypassed packet I/O and 5-tuple-based flow classification.
Gen features the dynamic scheduling of GPU threads to elastically scale VNF instances. It also supports runtime SFC modification by exploring features of the CUDA library. Also, it maintains a connection with a remote controller to orchestrate SFC execution.
Grus reduces processing latency through a set of data-path optimizations, including coordinated access to the PCIe bus between the CPU and GPU, fine-grained scheduling of consolidated VNFs, and dynamic batching. 
G-NET deploys network functions in VMs and offloads VNF processing to a GPU. It manipulates the GPU context to allow for spatial GPU sharing across manifold VNF kernels and leverages safe pointers to guarantee GPU memory isolation. A scheduling algorithm is designed to calculate the per-SFC cost and optimize the GPU resource sharing.


\section{Design Space} \label{sec:design}
In this section, we explore the design space, identify the critical design issues, and summarize different choices adopted by existing NFV platforms. Specific design choices for each platform are illustrated in Table~\ref{nfv-sum}. Note that, in general, there is no superior choice among the others; it is only a matter of use cases and application context.

\setlength\extrarowheight{1.1pt}
\begin{table}[!tb]
\begin{adjustwidth}{-1in}{-1in}
  \centering 
  \tiny	
  \begin{tabular}{ |c | c c | c | c c | c c | c | c c | c | c c | c c c |ccccc| }
 \multicolumn{1}{r}{} & \multicolumn{5}{c}{MANO plane} & \multicolumn{5}{c}{Service layer} &\multicolumn{11}{c}{NFV infrastructure} \\ \hline
  \multirow{3}{*}{Platform} & \multicolumn{2}{c}{High-level} & \multicolumn{1}{c}{PE} & \multicolumn{2}{c|}{State} & \multicolumn{2}{c}{Execution} & \multicolumn{1}{c}{TCP} & \multicolumn{2}{c|}{VNF} & \multicolumn{1}{c}{Packet} & \multicolumn{2}{c}{VNF} & \multicolumn{3}{c}{Virtualization} & \multicolumn{5}{c|}{Other} \\
  & \multicolumn{2}{c}{API} & \multicolumn{1}{c}{} & \multicolumn{2}{c|}{redistribution} & \multicolumn{2}{c}{model} & \multicolumn{1}{c}{} & \multicolumn{2}{c|}{I/O} & \multicolumn{1}{c}{I/O} & \multicolumn{2}{c}{connect} & \multicolumn{3}{c}{technique} & \multicolumn{5}{c|}{optimizations} \\ \cline{2-22}
   & \multicolumn{1}{c}{GPL} & DSL & & \multicolumn{1}{c}{SM} & MA & \multicolumn{1}{c}{RTC} & PL & & Poll & ITR & & VS & CI & VM & CT & PR & BA & ZC & PE & CO & HO\\ \hline
 OpenBox & $\surd$ && $\surd$ && $\surd$ & $\surd$ & &  & & & Kernel && & $\surd$ && &&&&& $\surd$\\  
 E2 & & $\surd$ & $\surd$ & & $\surd$ & $\surd$ && $\surd$ &&& DPDK & $\surd$ & & & & &&&& $\surd$&\\  
 SDNFV & & & $\surd$ & & & $\surd$ &&  & & $\surd$& DPDK & & $\surd$ & $\surd$ && & $\surd$ & $\surd$ & $\surd$ && $\surd$\\  
 Slick & $\surd$ && $\surd$ & & & $\surd$ && $\surd$ &&& Kernel && & && $\surd$ &&&&& \\  
 Eden & & $\surd$ & $\surd$ & & & && $\surd$ & & & Kernel & & $\surd$ & & & $\surd$ &&& $\surd$ && $\surd$ \\  
 MicroNF & & & $\surd$ && $\surd$ & & $\surd$ &  & & $\surd$ & DPDK & $\surd$ & & $\surd$ & $\surd$ & &&&&&\\  
 $\mu$NF & & $\surd$ &  && $\surd$ & & $\surd$ &  & $\surd$ && DPDK && $\surd$ & & $\surd$ & $\surd$ & $\surd$ & $\surd$ & & &\\  
 Metron & $\surd$ & & $\surd$ & $\surd$ & & $\surd$ & &  & & & DPDK & $\surd$& & && $\surd$ & & & & & $\surd$\\  
 Flurries & $\surd$ && && & && $\surd$ & $\surd$ & $\surd$ & DPDK & & & & $\surd$ & & $\surd$ & $\surd$ &&& $\surd$\\  
 MicroBoxes & $\surd$ && && & && $\surd$ &&& DPDK & & & & $\surd$ & & $\surd$ & $\surd$ &&& $\surd$\\  
 OpenNF & & &  & $\surd$ & & & & $\surd$ & & & Kernel & & & && &&&&&\\  
 NFVactor & $\surd$ & & & $\surd$ & & $\surd$ & & & $\surd$ & & DPDK & $\surd$ & & & $\surd$ & && $\surd$ & $\surd$ &&\\ 
 ClickNF & $\surd$ & &  & & & $\surd$ && $\surd$ & $\surd$ & & DPDK & & & & & $\surd$ & $\surd$ &$\surd$& $\surd$ & $\surd$ & $\surd$ \\  
 Flick & & $\surd$ &  & & $\surd$ & & $\surd$ & $\surd$ & & $\surd$ & DPDK & & & & & $\surd$ &&& $\surd$ && \\  
 NetStar & $\surd$ & &  & & $\surd$ & $\surd$ && $\surd$ & $\surd$ && DPDK & & $\surd$ & && $\surd$ &&& $\surd$ && \\  
 S6 & $\surd$ & &  & & $\surd$ & & $\surd$ & $\surd$ & & $\surd$ & DPDK & $\surd$ & & & $\surd$ & $\surd$ &&& $\surd$ && \\  
 StatelessNF & $\surd$ & &  & & $\surd$ & && $\surd$ & $\surd$ && DPDK & & & & $\surd$ & & $\surd$ & $\surd$ &&& \\  
  \hdashline
 \multirow{3}{*}{libVNF} & \multirow{3}{*}{$\surd$} & &  & & \multirow{3}{*}{$\surd$} & & & \multirow{3}{*}{$\surd$} & & & DPDK & \multirow{3}{*}{$\surd$} & & \multirow{3}{*}{$\surd$} & & \multirow{3}{*}{$\surd$} &  \multirow{3}{*}{$\surd$} & &  \multirow{3}{*}{$\surd$} &  \multirow{3}{*}{$\surd$} & \\
 & & &  & & & & & & & & netmap & & & & & &&&&&\\ 
 & & &  & & & & & & & & kernel & & & & & &&&&& \\  
 \hdashline
 NFMorph & & $\surd$ &  & & & & $\surd$ & $\surd$ & $\surd$ & & DPDK & & $\surd$ & & & $\surd$ & $\surd$ & $\surd$ & & $\surd$ & $\surd$\\  
 NetVM & $\surd$ & &  & & & $\surd$ & &  & & $\surd$ & DPDK & $\surd$ & & $\surd$ && & $\surd$ & $\surd$ &&& \\  
 OpenNetVM & $\surd$ &&  & & & $\surd$ & &  & $\surd$ && DPDK & $\surd$ & & & $\surd$ & $\surd$ & $\surd$ & $\surd$ &&& $\surd$ \\  
 NetML & $\surd$ &&  & & & $\surd$ & &  & $\surd$ && DPDK & $\surd$ & & & $\surd$ & $\surd$ & $\surd$ & $\surd$ &&& $\surd$ \\  
 NetBricks & $\surd$ & &  & & & $\surd$ & & $\surd$ & $\surd$ && DPDK & $\surd$ & & & & $\surd$ & $\surd$ & $\surd$ & $\surd$ && \\  
 ClickOS & $\surd$ &&  & & & $\surd$ & &  & & $\surd$ & netmap & $\surd$ & & $\surd$ & & & $\surd$ &&&& \\  
 HyperNF& & & & & & & & & && netmap & $\surd$ & & $\surd$ && & $\surd$ &&&& \\  
 IOVTee & & & & & & &&&& & DPDK & & $\surd$ & $\surd$ & $\surd$ & & & $\surd$ &&& \\  
 CHC & & & && $\surd$ & &&& $\surd$ & & VMA & & $\surd$ & & $\surd$ & &&& $\surd$ && $\surd$\\  
 NICA & $\surd$ & & & & & && $\surd$ & $\surd$ & & VMA && & $\surd$ & & & $\surd$ & $\surd$ && $\surd$ & $\surd$\\  
 Polycube & $\surd$ & & && & & & $\surd$ &&& eBPF && & $\surd$ & & &&&&&\\  
 NFP & & & && & & $\surd$ &&&& DPDK & & $\surd$ & & $\surd$ & & $\surd$ & $\surd$ & $\surd$ && \\  
 ParaBox & && && & &&&&& DPDK & $\surd$ & & & $\surd$ & &&& $\surd$ && \\  
 NFVNice & $\surd$ && && & && $\surd$ & & $\surd$ & DPDK & & $\surd$ & & $\surd$ & & $\surd$& $\surd$&&&$\surd$\\  
 PA-Flow & & & & && & $\surd$ & & & & DPDK & $\surd$ & & $\surd$ & $\surd$ & $\surd$ & $\surd$ & $\surd$ &&&\\  
 \hline
  \end{tabular}
\end{adjustwidth}
  \caption{Design choices for most of the existing NFV platforms. 
  }
  \label{nfv-sum}
\end{table}

\subsection{MANO plane}

\subsubsection*{High-level API}
Most existing NFV frameworks provide high-level APIs for specifying service policies or streamlining VNF development. These APIs can generally come as either {\bf Domain-Specific Language (DSL)} or {\bf General-Purpose Language (GPL)}.
GPLs such as C, C++, Java, and Python are mature programming languages capable of solving problems in multiple domains. They are shipped with multitudinous control primitives, miscellaneous data structures, and flexible operating patterns. Most existing NFV platforms are licensed under the GPL. For example, OpenNF relies on a collection of northbound C++ functions to develop control applications. NFVNice exposes a C library named ``libnf" to perform I/O operations asynchronously and to monitor the workload (e.g., arrival rate, processing time) for each VNF. OpenBox exposes its northbound Java API to network operators, allowing them to specify processing logic and subscribe to specific events. Slick provides a programming abstraction that allows developers to write custom VNFs and specify traffic steering policy in Python.
ClickNF provides a standard socket API and a zero-copy interface, both implemented as C++ functions, for interacting with its transport layer.
S6 exposes a programming API to manipulate states across the shared object space.
NICA introduces the ``ikernel" programming abstraction over TCP/UDP sockets to facilitate user-space VNF development. 
GPUNFV exposes a CUDA-based API to assist per-flow state abstraction and construct GPU kernel code. FlowShader and G-NET provide a CUDA-based API to develop VNFs compatible with both GPU and CPU semantics.
Compared to GPLs, DSLs provide higher-level, optimized abstractions for specific problems and usually operate in environments with limited operational patterns and restricted resource usage. 
Several platforms incorporate DSL for network service development. For example,
Alim et al.~\cite{alim2016flick} propose flick language that supports parallel execution and safe resource sharing. In addition to basic primitives such as event handling and common data types, Flick can deserialize input packets into application-specific data types and vice versa, bringing application semantics into VNF development. 
E2 allows network operators to specify SFCs using a policy language, and SONATA's development toolchain allows developers to specify network services using DSL. On the Eden platform, network operators specify service policies in the F\# language, which makes the VNF safety-checking process straightforward. 
Rubik~\cite{li2021rubik} exemplifies a “composition-first” high-level API: instead of only exposing
primitives for packet handling, it structures NF programs as reusable pipeline components and supports systematic transformations of the composed service.


\subsubsection*{Placement}
During the VNF placement phase, existing NFV platforms typically use a Placement Engine (PE) that performs a set of \textbf{pre-processing} operations to merge and shorten SFCs before actually \textbf{placing} them at the anticipated NFVI PoPs with specific objectives. For instance, based on the specified network service description and infrastructural specification, CoMb seeks to consolidate each SFC on a single NFVI-PoP by solving an optimization model.
OpenBox relies on a graph-merging algorithm to optimally deploy VNFs to the user-specified NFVI-PoPs. Through the OpenBox northbound interface, developers can specify their VNF as a processing graph along with the intended deployment domain or NFVI-PoP. The OpenBox controller parses the graphs intended for the same location and merges them into a single graph without violating their processing logic.
Slick provides a holistic solution for VNF placement and traffic steering. The Slick controller employs an inflation heuristic to consolidate VNFs with minimum cost and uses a placement algorithm to deploy the consolidated VNFs. Traffic steering rules are also configured on each switch to realize the intended routes and processing sequence for each flow.
E2 employs similar approaches to merge and synthesize multiple service graphs to reduce processing redundancy. Then it models VNF instance placement as a graph partition problem over the COTS servers and employs a modified Kernighan-Lin algorithm to minimize the inter-server traffic.
SDNFV formulates the service placement problem as a mixed-integer linear program (MILP) to maximize resource utilization. Then, the authors develop a heuristic algorithm to place VNFs and configure related traffic routes based on the service graph specified by the network operator.
Metron uses SNF to optimize its input processing graph and construct a synthesized graph, which is subsequently split into a stateful subgraph and a stateless subgraph. The stateful graph is deployed on COTS servers selected by Metron's server selection scheme. The stateless graph is then offloaded to network elements based on the deployment locations of the stateful graph.
MicroNF performs dependency analysis on elements rather than VNFs and reconstructs the service graph to reduce redundant processing and improve resource efficiency. It subsequently places the modularized SFCs on the COTS servers by resolving a 0-1 integer programming problem to minimize the inter-VM overhead.
The $\mu$NF orchestrator also constructs an optimal forwarding graph by consolidating the same types of VNFs, but the objectives based on which the VNFs are placed across COTS servers are not indicated by the authors.


\subsubsection*{State redistribution}

With the proliferation of stateful VNFs, it is hence critical to maintain consistent processing states during instance scaling. However, it is extremely challenging to simultaneously satisfy all the state management requirements, e.g., flow affinity preservation, timely state synchronization, correct processing, minimal service interruption, etc, especially upon VNF scaling or instance failover. Strategies adopted by existing NFV platforms can be generally classified into \textbf{State Migration} (SM) and \textbf{Migration Avoidance} (MA).
Among the platforms that adopt state migration, OpenNF provides the most robust scheme by allowing control applications to move, copy, and share states of varying granularity between two VNF replicas. In particular, it implements state-movement and copying operations to facilitate state migration. The move operation applies cautious coordination between source/destination instances and the last shared on-path OpenFlow switch to achieve lossless, order-preserving state migration. The copy operation allows state clones in an eventually consistent manner, while the share operation guarantees strong consistency by instructing the controller to capture all events from the last shared switch, send them to the corresponding VNF instances for processing, and apply state updates sequentially in the global scope. MicroNF, UNO, and OpenBox also advocate this solution for state coordination.
Split/Merge also relies on the SDN mechanism to migrate partitioned states and flows across multiple VNF replicas. Upon a scaling decision, it first instructs SDN to suspend traffic for all replicas, then transfers relevant states across the replicas and configures routes for the affected flows, and finally resumes traffic. This approach preserves flow affinity, but the in-transit packets for the affected flow are lost.
Metron migrates states by traffic classes. When an SFC is overloaded, Metron splits its traffic classes into two groups and simply duplicates only the states of the migrated group to the new SFC instance. Compared with OpenNF, this scheme may lead to runtime inconsistencies and memory waste.
TFM migrates states between two VNFs by instantiating a TFM box on each VNF. Then, the TFM controller transmits flow states to the destination VNF and updates the routing tables of the in-path switches to redirect incoming packets of migrating flows to the destination VNF. The TFM boxes at both VNFs buffer all in-transit packets, which are fed to the destination VNF in the order received after state migration.
In contrast, other platforms adopt a migration-avoidance strategy to reduce state-migration overhead. 
For instance, upon scaling out, E2 splits incoming traffic based on its flow identifiers (e.g., 5-tuple) on the original VNF and identifies new flows that need to be directed to the newly instantiated VNFs. Then the E2 manager configures routing tables of the in-path software and hardware switches to steer new flows from the original VNF to new instances. Existing flows will continue to be served by the original VNF until termination. Although the traffic direction phase may incur additional delay, the authors believe this overhead is transient, and this approach outperforms state-migration approaches.
MicroNF employs a "Push-Aside" scaling strategy in which the overloaded element kicks its upstream/downstream element to a neighboring VM to free resources. Therefore, instead of migrating states, MicroNF moves VNF elements, avoiding state migration. But it is unclear whether the element migration phase can cause any data loss or inconsistency.
The most common migration-avoidance strategy is to externalize the processing state.
StatelessNF, CHC, and libVNF store VNF processing states in high-speed external data stores to avoid migration costs.
In particular, libVNF maintains key-value data stores at multiple levels: a local data store shared by all threads of a single VNF, a global store shared across multiple VNF replicas, and a local cache for recently accessed data from the global store.
Disaggregating Stateful Network Functions~\cite{bansal2023disaggregating} exemplifies an architectural choice for state redistribution: instead of moving state with compute, the platform can decouple state from packet processing and manage it as a shared substrate to simplify scaling and recovery.

%
%


\subsection{VNF plane}\label{sec:nf-design}
To implement efficient VNFs, NFV platforms are required to consider the following design choices:

\subsubsection*{Execution model}
\begin{figure}[!tb]
\centering
\includegraphics[width=0.5\textwidth]{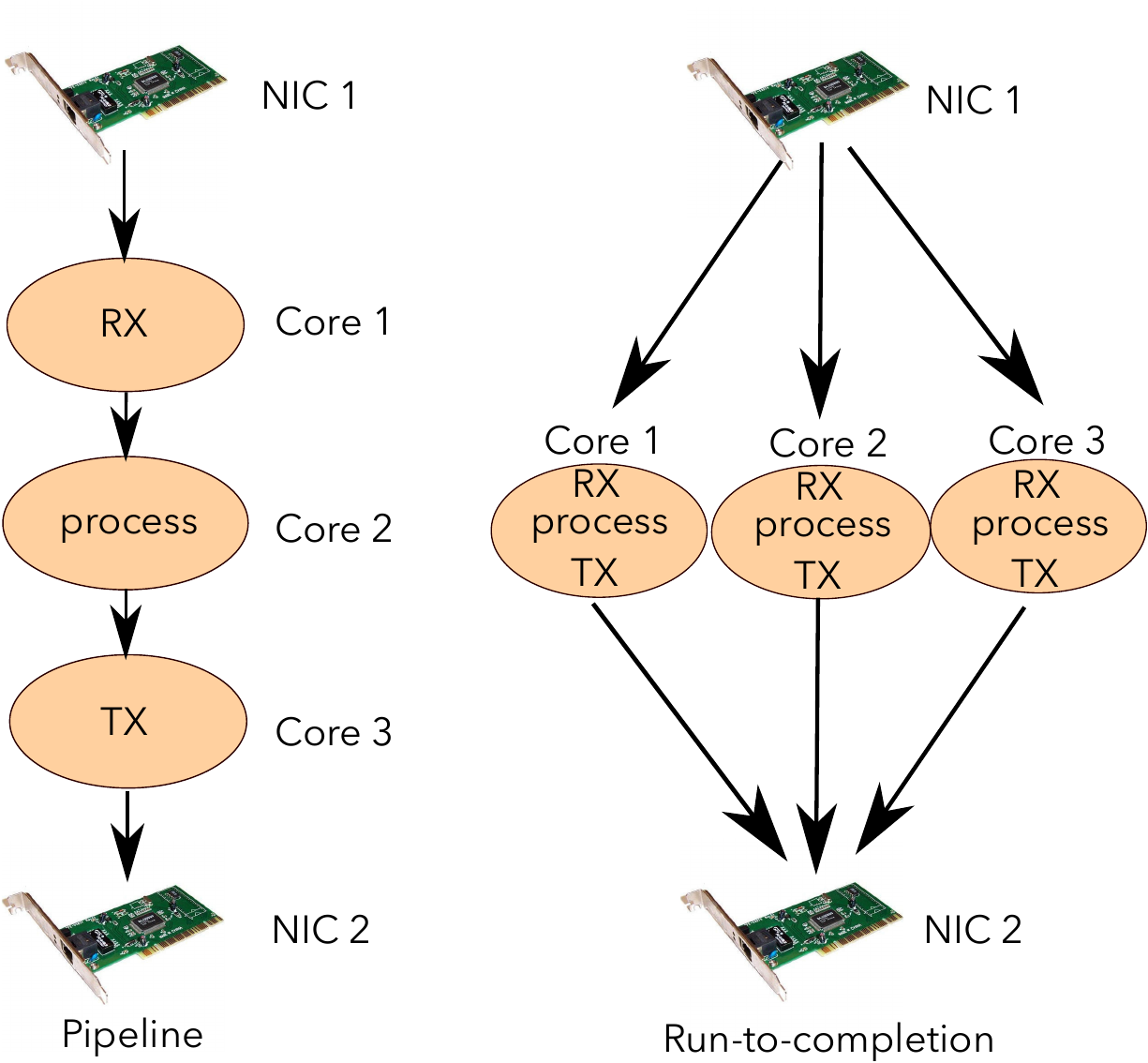}
\caption{Execution models of a simple VNF instance sitting between two physical NICs. It consists of three elements: RX, processing, and TX. }
\label{fig:exe}
\end{figure}

In the NFV domain, the VNF (or SFC) execution model can be classified into \textbf{Run-To-Completion (RTC)} and \textbf{Pipeline (PL)}. In the RTC model, all the VNFs of a given SFC run on a single thread, whereas in the pipeline model, each VNF instance is pinned to a separate thread, as illustrated in Fig.~\ref{fig:exe}. The performance of either model depends heavily on processing complexity and input workload, resulting in different cache and memory access patterns. In general, the RTC model achieves higher throughput and lower latency when executing simple VNFs and/or short SFCs, since it avoids inter-core transfer overhead~\cite{katsikas2018metron}. It may also require fewer worker cores than the pipeline model. However, the edge of RTC fades with complex VNFs and/or lengthy SFCs, in which cases a large amount of data/instructions and context switches might precipitate persistent L1/L2 cache misses. VNFs running in the RTC model are also more difficult to scale individually based on input load. As a result, there is no single superior solution, and distinct NFV platforms typically make choices based on their specific architectures. In \cite{zheng2019closer}, the authors propose a hybrid execution model that leverages these two models while avoiding their shortcomings. 
Some platforms adopt the RTC model mainly because they only execute lightweight VNFs or trimmed SFCs in software.
For instance, NetBricks, ClickNF, ClickOS, NetStar, NFVactor, and SafeBricks execute VNFs as processes in the RTC model to avoid inter-core transfer overhead.
Metron offloads part of its SFCs to in-path hardware, executing the trimmed SFC tasks in the RTC model on COTS servers. CoMb executes an entire VNF or SFC in a single core to avoid inter-core synchronization overhead. 
GPU-based solutions also adopt the RTC model because the pipeline model is inherently inefficient on GPU architectures~\cite {zheng2018gen}. Consequently, FlowShader, GPUNFV, and Gen each embed a single integral SFC within a single GPU thread.
Other platforms employ the pipeline model. For example, 
$\mu$NF and MicroNF decompose VNFs or SFCs into modular, fine-grained, loosely-coupled packet processing tasks (or elements) so that resources can be precisely allocated to scale individual elements. 
Flick enables the pipelined execution of individual tasks within its VNFs.
PA-Flow is designed for SFCs distributed across NFVI-PoPs and running in pipeline mode.
The poll-mode VNFs scheduled by UN$i$S also run in pipelines constructed with DPDK ring buffer. ShieldBox also adopts DPDK ring buffers to VNFs across multiple SGX enclaves.
StatelessNF executes VNFs in pipelines. Each pipeline consists of a polling thread, a lockless queue, and a processing thread.
Maestro~\cite{pereira2024maestro} exemplifies platforms that treat parallelization as an execution-model concern rather than an NF developer burden: the platform assists in extracting safe parallelism while preserving the NF’s intended semantics.
NFOS~\cite{yan2024transparent} highlights a different execution-model choice: instead of forcing NFs to be written in a parallel style, the platform can provide transparent scaling support and manage concurrency as a substrate responsibility.


\subsubsection*{TCP functionality}
As stateful VNFs have become an important building block in the NFV ecosystem, it is worth noting existing platforms that implement or integrate a TCP/IP stack to support stateful VNFs at layer 4 or beyond.
ClickNF is equipped with a full-fledged modular TCP stack to facilitate the end-host application development.
Microboxes include a modular, customizable TCP stack that can be shared among a group of VNFs, eliminating redundant processing. NICA even implements a simplified TCP stack in SmartNICs to enrich its in-path processing features.
xOMB stack also implements simple functions, such as terminating TCP connections.
Instead of developing TCP functionalities from scratch, some platforms choose to directly incorporate third-party solutions. For example, Flick integrates the high-speed kernel-bypassing mTCP stack~\cite{jeong2014mtcp} to implement transport-layer VNFs, whereas NetStar directly employs a third-party user-space TCP stack with future/promise abstraction.
libVNF is designed to be generic by integrating both the standard networking stack and mTCP.
Moreover, all NFV platforms that use the kernel TCP/IP stack are granted TCP functionality by default. 
In particular, Polycube directly cooperates with the kernel TCP/IP stack to build complex SFCs.


\subsubsection*{VNF I/O}
Just like the execution model, there are also two alternative means for VNFs to perform packet I/O, namely \textbf{Polling} and \textbf{Interrupt (ITR)}. VNFs running in polling mode repeatedly query the NICs or upstream VNFs for data, which typically yields better performance at the cost of wasted CPU cycles and increased energy consumption. Interrupt-based I/O usually does not entail wasted resources but incurs performance losses due to interrupt propagation delays and cache line warm-up. 
In existing NFV platforms, UNO, CHC, NFVactor, and StatelessNF execute VNFs in poll-mode to enhance performance. UNiS is tailored to schedule poll-mode VNFs by manipulating the Linux real-time scheduler. 
Flick, ClickNF, NFVNice, xOMB, and libVNF execute VNFs in interrupt mode.


\subsubsection*{Secure execution}
At present, it is increasingly common for VNFs to be delegated for execution in untrusted environments such as public or third-party clouds. Consequently, both traffic data and VNF information are exposed to different cyber attacks. 
Existing NFV platforms generally secure the execution of VNFs with either \textbf{Encryption approach} or \textbf{Shield Execution}. Platforms that adopt an encryption approach typically use various cryptographic schemes to enable VNFs to operate directly on encrypted network traffic. 
On the other hand, platforms adopting shield execution commonly run VNFs in private memory regions called enclaves. The contents of an enclave are strictly protected and cannot be accessed by any external process. Compared to shield execution, encryption approaches generally incur higher overhead due to the complex cryptographic computations and support a limited set of functionalities. The only advantage is that they are not reliant on trusted hardware such as Intel SGX.


\subsection{NFVI}

\subsubsection*{Virtualization technique} \label{sec:virt}
As the central point of any NFV platform, existing implementations typically deploy network functions inside \textbf{Virtual Machines (VMs)}, \textbf{Containers (CT)}, or simply as bare-metal \textbf{Processes (PR)}. 
\begin{figure}[!tb]
\begin{center}
\includegraphics[width=0.68\textwidth]{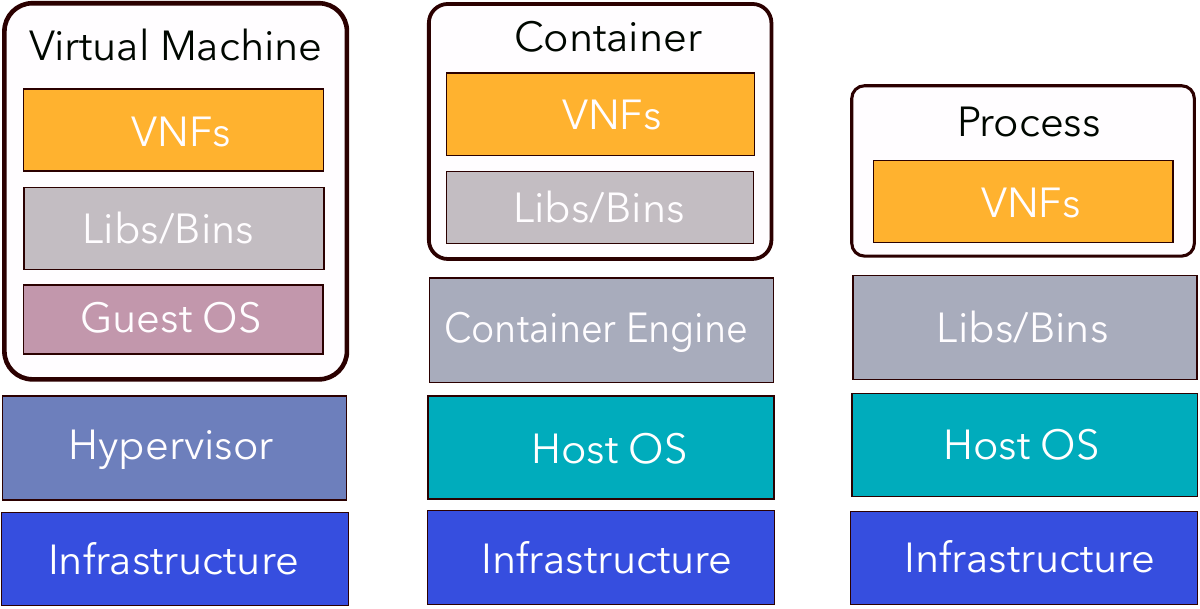}
\caption{State-of-the-art virtualization techniques}
\label{fig:virt}
\end{center}
\end{figure}
As illustrated in Fig.~\ref{fig:virt}, they realize virtualization at different layers in the commodity hardware infrastructure, and therefore present different degrees of isolation and resource requirements. 
VM is a hardware-level virtualization technique that relies on the Virtual Machine Monitor (VMM) or hypervisor, a firmware that offers a virtual platform to accommodate a variety of guest operating systems, on which different applications or VNFs execute\footnote{Note that there is another kind of hypervisor that operates at OS-level (namely hosted hypervisor), we exclude from our discussion as they are rarely used by existing NFV platforms.}. Hypervisors also supervise and coordinate VM instances to enable efficient hardware resource sharing across them.
As the most traditional approach for virtualization, VM-based NFV platforms are commonplace. For example, NetVM, NICA, and SDNFV exploit KVM-based VMs, Split/Merge, HyperNF, and FlowOS build VNFs inside Xen-based VMs.
However, traditional VMs commonly incur heavy resource demands, huge memory footprints, and high migration costs. To address these issues, ClickOS and CliMBOS adopt unikernel VMs, which are minimalist, small, agile, and fast to boot.
The advent of containerization techniques, such as LXC and Docker, bestows another option. Compared to VM, containerization is an OS-level virtualization technique with a much smaller memory footprint and shorter instantiation time, much higher deployment density (up to thousands per server), and lower redistribution costs. Nonetheless, containers are not completely insulated from the host OS; therefore, they cannot provide the same level of isolation and security as VMs. In terms of performance, both can sustain line-rate processing by leveraging specific I/O paths and networking techniques~\cite{zhang2019comparing,zhang2021performance,zhang2019benchmarking}. Currently, many NFV platforms adopt containerized VNFs. For instance, OpenNetVM, Flurries, MicroBoxes, NFVNice, MVMP, NFP, ParaBox, MVMP, statelessNF, NFVactor, and GNFC execute their VNFs inside Docker containers. CHC and Iron run VNFs in LXC containers.
Aside from VMs and containers, NFV platforms deploy VNFs as processes or threads to trade off isolation for performance. These platforms usually assume trusted VNFs and infrastructure.
For example, NetBricks, ClickNF, libVNF, and SafeLib execute VNFs as processes. OPNFV barometer is designed to profile the performance of the VNFs running as DPDK processes. SplitBox executes VNFs as processes inside FastClick. GPUNFV, Gen, FlowShader, Grus, and G-NET execute VNFs or SFCs as GPU threads.
Notably, some platforms offer multiple-choice options or introduce hybrid solutions. For example, NFV-VIPP and PF-Flow can be integrated into any DPDK-accelerated VNFs running inside VMs, containers, or as bare-metal processes. VNFs in OpenNetVM and $\mu$NF can be deployed either as processes or inside containers. MicroNF even runs containers inside VMs, probably to improve security.

\textbf{MicroVMs and secure container runtimes.}
In the last few years, the dichotomy between ``VMs for strong isolation'' and ``containers for lightweight deployability'' has been increasingly blurred.
A representative example is the microVM line of work, which preserves the VM boundary while aggressively minimizing the virtual machine monitor (VMM) and the exposed device model to approach container-like density and startup times.
Firecracker is a prominent instance, designed for multi-tenant serverless workloads where fast elastic scaling must coexist with strong isolation \cite{agache2020firecracker}.
In parallel, secure container runtimes revisit the container/guest boundary from different angles: some pursue a userspace-kernel sandbox that interposes on system calls and I/O paths (e.g., gVisor) \cite{young2019truecost}, while others decouple or refactor guest-kernel functionality to reduce the user--host interface and the trusted computing base, while preserving compatibility and performance \cite{manakkal2025liteshield,shi2025hardware}.
For NFV, these sandboxes are particularly relevant when VNFs are sourced from third parties or when multi-tenancy and rapid scaling are first-order design constraints, making ``containers inside VMs'' a principled point in the isolation--deployability design space rather than an ad hoc hybrid choice.

\textbf{Cold-start versus steady-state tradeoffs.}
As NFV increasingly adopts cloud-native control planes and on-demand scaling policies, the cost profile of a sandbox should be measured not only by steady-state packet-processing throughput but also by cold-start latency, memory footprint, and the operational overhead of frequent instantiation.
Catalyzer shows that aggressive specialization and initialization-less boot mechanisms can push startups toward the sub-millisecond regime for serverless-style deployments \cite{du2020catalyzer}.
This perspective complements classic NFV discussions on migration and placement cost by emphasizing that, under bursty workloads, startup overhead can dominate the time-to-serve budget.

\textbf{Fast bring-up of passthrough I/O for secure sandboxes.}
High-performance VNFs frequently rely on kernel-bypassing and passthrough paths, yet such paths may incur non-trivial setup overhead that directly impacts scaling speed.
Recent measurement studies indicate that, for secure containers, the \emph{network startup path} (e.g., CNI/plugin orchestration and device setup) can become a key bottleneck, especially under high-volume concurrent invocations \cite{liu2024imcstartup}.
Motivated by this, FastIOV argues that the startup performance of SR-IOV-enabled secure containers can be substantially improved, making passthrough networking more suitable for fast-scaling secure deployments \cite{liu2025fastiov}.
Overall, these results suggest that NFV platforms should treat ``I/O bring-up cost'' as a first-class metric when selecting an isolation mechanism, rather than assuming that passthrough performance automatically implies good elasticity.


\subsubsection*{VNF interconnects} \label{sec:int}
NFV platforms achieve cost efficiency by consolidating multiple VNF instances on a single COTS server. These VNFs are then purposely concatenated to form specific SFCs. Existing NFV platforms interconnect local VNFs using either state-of-the-art \textbf{Virtual Switches (VS)} or employ {\bf Custom Implementations (CI)}.
Software virtual switches are widely used as the NFV data plane for efficient traffic steering. For instance, E2, ParaBox, and NFVactor reuse BESS~\cite{han2015softnic} as the data plane to interconnect VNFs or runtimes, while ClickOS and HyperNF extend the VALE switch~\cite{rizzo2012vale}. CoMb customizes the Click Modular Router~\cite{kohler2000click} to classify and forward packets between VNFs of the same service chain.
Metron, MiddleClick, SCC, and SplitBox leverage FastClick~\cite{barbette2015fast} to ferry packets between VNFs and the network. Split/Merge, TFM, and MicroNF employ Open vSwitch (OVS)~\cite{pfaff2015design} for VM-networking, while UNO and PA-Flow employ OVS-DPDK~\cite{ovs-dpdk}.
Note that UNO steers packets at both the COTS server and SmartNIC. 
More details about the performance of the aforementioned software switches can be found in the benchmark study conducted by Zhang et al.~\cite{zhang2019comparing, zhang2021performance, zhang2019benchmarking}.
Rather than adopting third-party solutions, G-NET uses a bespoke software switch to route packets between VNFs and physical NICs.
OVS-centric acceleration~\cite{rashelbach2022scaling} shows that the interconnection substrate (e.g., a virtual switch) is itself a design lever: optimizing the shared vSwitch datapath can raise end-to-end NFV throughput without modifying individual VNFs.


\subsubsection*{Packet I/O frameworks} \label{sec:io}

\begin{figure}[!tb]
\centering
\includegraphics[width=0.65\textwidth]{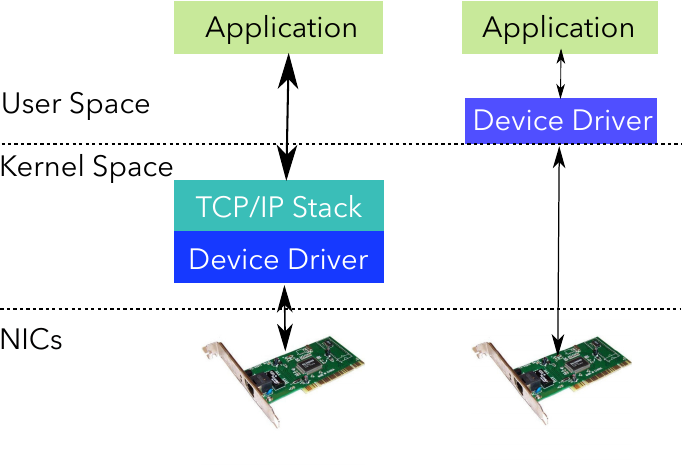}
\caption{Packet I/O: Kernel-based vs. Kernel-bypassing}
\label{fig:io}
\end{figure}

Packet I/O frameworks can be classified into \textbf{Kernel-based} and \textbf{Kernel-bypassing} approaches, as illustrated in Fig.~\ref{fig:io}. Traditionally, network applications rely on the general-purpose OS kernel stack for packet I/O. For instance, Eden, FlowOS, and NetContainer employ kernel-based I/O to seamlessly utilize the rich features of the kernel stack.
However, the OS kernel imposes non-negligible overhead on the packet data path, making software applications fail to sustain high-speed processing~\cite{barbette2015fast}. To overcome this bottleneck, kernel-bypassing frameworks such as DPDK~\cite{dpdk} and netmap~\cite{rizzo2012netmap} are proposed. They commonly include data-path optimization routines such as zero-copy delivery to avoid staging packets in kernel space, preallocated packet buffers to avoid runtime memory allocation, and batch processing to amortize the overhead of accessing hardware.
netmap is compatible with the standard kernel stack as it still incorporates system calls for data validation and interrupt-based packet reception. DPDK employs complete kernel bypassing and dispatches poll-mode drivers to enhance performance. DPDK further exposes a rich collection of APIs and primitives to simplify application development. 
As shown in Tab.~\ref{nfv-sum}, most existing NFV platforms use DPDK for packet I/O from physical NICs. For example, Flick, ClickNF, NetStar, NetVM, OpenNetVM, CoNFV, OPNFV barometer, NFV-VIPP, SafeLib, SplitBox, UNO, GPUNFV, Gen, Grus, G-NET all leverage DPDK for packet I/O. netmap is used by ClickOS and HyperNF. NICA and CHC rely on the Mellanox Message Accelerator (VMA)~\cite{vma}, another kernel-bypass packet I/O framework that provides standard POSIX socket APIs and a user-space networking library.
Note that even if the traditional kernel-based approach fails to achieve comparable performance to kernel-bypassing stacks, it can still be useful when the VNFs are not I/O-intensive or when the cost of setting up a kernel-bypassing stack becomes too high.
Furthermore, the Polycube platform adopts a kernel-based high-speed packet I/O framework, eBPF, to achieve multi-gigabyte processing.
libVNF is a general-purpose platform that integrates the kernel, DPDK, and netmap, allowing users to choose the components that best suit their needs.


\textbf{In-kernel fast path with eBPF/XDP.}
While kernel-bypassing frameworks (e.g., DPDK) remain the dominant choice for high-speed packet I/O in NFV platforms, post-2020 systems increasingly revisit the kernel as an execution substrate for network functions.
In particular, eBPF programs can be attached to multiple kernel hook points to intercept packets and enforce policies without fully bypassing the kernel stack.
Compared with kernel bypassing, the in-kernel approach offers attractive operational properties: it reuses the existing kernel networking stack, reduces the need for bespoke drivers, and supports dynamic updates subject to the safety constraints enforced by the kernel verifier.
Polycube~\cite{miano2019service} is a representative platform that uses eBPF to achieve multi-gigabyte packet processing while still integrating with the kernel stack to build complex service chains. 
More recently,library-style designs aim to make in-kernel network function development more modular and performance-oriented.
For example, eNetSTL~ \cite{YangEuroSys25eNetSTL} advocates an in-kernel library abstraction to support high-performance eBPF-based network functions, reflecting a trend toward “standardized building blocks” rather than ad-hoc, monolithic eBPF programs.

\textbf{Lifecycle, composition, and testing at fleet scale.}
A key obstacle to adopting in-kernel network functions in production is not only raw performance but also the end-to-end lifecycle management of a large number of coexisting eBPF programs: composition across hook points, safe rollout, configuration management, and rollback.
NetEdit~\cite{BensonSIGCOMM24NetEdit} provides an orchestration platform that explicitly targets these issues, offering (i) unified abstractions across diverse hook-points, (ii) a configuration language that decouples policies from programs, (iii) explicit object lifecycle management, and (iv) extensive testing methods.
NetEdit reports multi-year production deployment and substantial performance impact, indicating that eBPF-based network function orchestration has matured from a “programming technique” into a platform-level capability. 

\textbf{Practical concerns: interface evolution and portability.}
Despite the above progress, in-kernel extensibility raises new engineering risks that are less visible in user-space packet processing stacks.
In particular, kernel hook-points, helpers, and program semantics evolve with kernel versions, and production deployments must cope with dependency and compatibility issues across heterogeneous fleets.
Zhang et al.~\cite{ZhongEuroSys25UnstableEBPF} highlight these unstable foundations of eBPF-based kernel extensions, motivating systematic compatibility engineering and regression testing as first-class platform concerns.


\subsection{Other design choices}\label{sec:opt}
As discussed in our previous work~\cite{linguaglossa2019survey}, there is a large assortment of acceleration techniques for high-speed packet processing. In this section, we select the most commonly used techniques and enumerate their adoption across existing NFV platforms. The optimizations we consider include zero-copy, batching, memory pre-allocation, instruction/data prefetching, parallel execution, CPU cache optimization, and computation offloading. Although these optimizations are commonly applied by the preceding packet I/O techniques, we discuss their applications in other parts of the NFV architecture.


\subsubsection*{Zero Copy (ZC)}
In the high-speed packet processing domain, runtime memory copy is an expensive operation that typically incurs prohibitive overhead. For performance, many existing NFV platforms deliver packets across VNFs or memory boundaries in a zero-copy manner, copying only their associated packet descriptors. 
For example, $\mu$NF implements a zero-copy port abstraction that only exchanges packet addresses instead of copying full packets between VNFs.
The TCP stack in ClickNF exposes zero-copy interfaces for interacting with the user-space VNF.
NetVM and OpenNetVM implement a zero-copy packet mechanism through shared memory.
NICA leverages ring buffers for zero-copy message exchange between the F-NIC units and the user-space VNFs.
GPUNFV and NetML achieve zero-copy packet delivery across the CPU-GPU boundary using CUDA's page-locked memory. 
G-NET's switch also employs a zero-copy design.

\subsubsection*{Batching (BA)}
In high-speed packet processing frameworks, I/O batching is widely used to amortize the overhead of accessing the physical NIC over multiple packets. This technique is also employed by some NFV platforms to enhance performance. For example, NFVNice and EdgeMiner batch the I/O interrupts to amortize VNF wakeup overhead.
SCC handles VNF I/O system calls in dynamic batches to reduce context-switch overhead.
The VNFs on the $\mu$NF platform perform packet I/O in batches through the intermediate ring buffers.
The TCP stack of ClickNF exchanges packets with the user-space VNFs in batches.
StatelessNF aggregates multiple read/write requests to the data store into a single request to amortize the overhead of remote procedure call (RPC).
SafeBricks implements an in-enclave module for batched packet I/O from/to the host.
LightBox adopts packet batching to amortize the system call overhead.
GPUNFV, Grus, FlowShader, G-NET, and Gen deliver packets between CPU and GPU in dynamic batches.

\subsubsection*{Pre-allocation}
Memory allocation at runtime remains an expensive operation. In addition to pre-allocated packet buffers and descriptors used by some packet I/O techniques, existing NFV platforms usually pre-allocate a dedicated memory region to stage and reuse other relevant packet-processing data structures. For example, libVNF pre-allocates memory pools for its per-core, persistent request objects, and lock-free packet buffers. Flick pre-allocates its task graphs and queues. 
S6 pre-allocates a pool of cooperative, user-space per-flow micro-threads to avoid the dynamic thread-creation/deletion overhead.
ShieldBox pre-allocates packet descriptor memory. LightBox pre-allocates state management data structures.

\subsubsection*{Parallel Execution (PE)}
To take advantage of multicore CPUs, many platforms explore parallelization. 
$\mu$NF performs a dependency analysis on its forwarding graphs to identify parallelize VNFs. Consecutive VNFs are deemed parallelizable if they perform read-only operations or update disjoint packet regions. These VNFs are then assigned independent CPU cores to process packets. 
A reference counter is attached as metadata to avoid out-of-order operations from downstream VNFs.
Likewise, SDNFV allows multiple VNFs to access a packet in parallel using a reference counter embedded in the packet descriptor. Eden exposes a concurrency model that creates consistent state copies for multiple VNFs to execute in parallel in the Eden enclave. CoMb allocates an independent shim layer to each SFC to enable parallel execution of multiple SFCs.
Flick instantiates a new task graph for each new connection and schedules the tasks of these graphs onto multiple worker cores in parallel.
ClickNF uses Receive-Side Scaling (RSS) on physical NICs to distribute incoming packets across multiple cores, with flow-level affinity guaranteed.
NetStar builds VNFs using a share-nothing thread model and distributes incoming packets across threads for parallel multicore processing.
libVNF is built with VNF multicore scalability and uses per-core data structures to avoid inter-core communication, which can hamper VNF multicore scalability.

\subsubsection*{Cache Optimization (CO)}
Modern CPUs are equipped with hierarchical caches between their cores and the main memory. Cache misses result in additional accesses to other cache levels or to main memory, significantly slowing processing speed. Many existing NFV platforms are aware of this issue and explore opportunities for cache optimization.
NetContainer aims to exploit cache locality at inter-flow and intra-flow levels for NFV workloads and leverages the page coloring technique to aggregate buffer pages into separate cache regions to avoid cache contention.
ResQ exploits Intel Cache Allocation Technology, along with corresponding buffer sizing, to eliminate last-level cache invalidation while ensuring performance isolation. LightBox adapts cache-line protection techniques to reduce cache miss rates.
$\mu$NF performs cache-line pre-fetching in batches to increase the cache hit rate.
Some platforms also cache-optimize their critical internal data structures. For example,
the request objects of libVNF are cache-optimized, and all the per-core data structures of ClickNF are cache-aligned.

\subsubsection*{Hardware Offloading (HO): SmartNIC/DPU and Programmable Gateway Era}
\label{sec:ho-2025}
Hardware computation offloading is widely adopted by existing NFV platforms to alleviate the pressure of COTS servers. Potential resources to offload computing tasks, including GPU, smartNICs, in-path programmable network equipment, or other specialized accelerators.
The E2 manager maintains a connection with the OpenFlow switch and cognitively offloads simple VNFs to unburden the servers. 
Likewise, Metron offloads stateless operations to the in-path programmable NICs and switches. 
OpenBox and Eden also support hardware implementation of their forwarding plane to accelerate processing.
OpenNetVM and OpenANFV incorporate programmable NICs or FPGAs for computation offloading.
ClickNF explores common NIC features to perform TCP/IP checksum offloading, TCP segmentation offloading (TSO), and large receive offloading (LRO).
GPUNFV, Gen, FlowShader, Grus, NetML, and G-NET all achieve performance gains by offloading at least some of their computation to the GPU.
SmartNICs are commonly equipped with programmable, multi-core processors and an integrated operating system, making them ideal for executing computation tasks.
UNO and NFMorph explore smartNICs to offload VNFs, forwarding rules, flow tables, and crypto/compression operations.
NICA leverages FPGA inline processing on smartNICs to accelerate data-plane processing. The ``ikernel" programming abstraction of NICA grants user-space VNFs direct control over the computations in SmartNICs.
CHC allows VNFs to offload operations to the external state store to speed up shared state updates.
Another flavor of hardware computation offloading comes with reusing a computation result.
For instance, OpenNetVM and NFMorph reuse the NIC's Receive-Side Scaling (RSS) hash value for traffic classification at a later stage. SDNFV caches flow table lookup results in packet descriptors for reuse by the VNFs.

Entering the 2020s, hardware offloading has evolved from an optional acceleration technique into a
\emph{platform design choice} that determines (i) where packet-processing logic executes, (ii) how state is partitioned, and (iii) what isolation and upgrade mechanisms are feasible. Beyond traditional FPGA- and GPU-based accelerators, modern deployments rely on two increasingly common substrates: \emph{(a) programmable gateways} built on switch ASICs and FPGAs, and \emph{(b) SmartNICs/DPUs} that host programmable packet-processing pipelines and on-board compute.

\paragraph{Programmable gateway platforms.}
Cloud gateways are no longer purely software switches running on x86; several systems demonstrate production-grade gateways in which the critical fast path is implemented in programmable hardware, complemented by software for elasticity and complex control. Sailfish~\cite{pan2021sailfish} is a programmable-switch gateway that targets multi-tenant, multi-service cloud ingress/egress at very high throughput, while preserving operational flexibility. LuoShen~\cite{pan2024luoshen} extends this direction towards a \emph{hyper-converged programmable gateway} for multi-tenant edge clouds by
co-designing a programmable data plane with a host-side control plane. More recently, the gateway substrate itself has diversified: Albatross~\cite{lu2025albatross} reports a containerized gateway platform that leverages FPGA-accelerated packet-level load balancing, illustrating a broader trend towards \emph{heterogeneous gateway pipelines} when pure switch-ASIC resources or evolution constraints become bottlenecks.
Nezha~\cite{li2025nezha} illustrates a modern HO choice: once parts of the vSwitch move into
SmartNICs, the platform must also decide how to allocate and share the limited on-NIC resources to avoid creating new contention bottlenecks.
Sirius~\cite{gao2024sirius} exemplifies HO choices at the chain level. It partitions functionality between programmable dataplanes and software, effectively mapping SFC logic to heterogeneous execution substrates.
SyNAPSE~\cite{pereira2022automatic} supports a different HO choice: instead of hand-building offloaded variants, it relies on synthesis/generation to produce accelerator implementations from reusable components, trading manual optimization effort for systematic exploration.

\paragraph{SmartNIC/DPU offloading toolchains.}
While SmartNICs/DPUs offer a natural place to offload network functions (NFs), achieving speedups is not automatic: performance depends on the NF's compute/memory profile, state layout, and the SmartNIC's micro-architecture. Clara~\cite{qiu2021clara} argues that developers need systematic guidance for offloading; it analyzes an NF and suggests porting strategies that can improve
offloaded performance. At the same time, SmartNIC programmability (e.g., P4 pipelines) introduces its
own optimization space; Pipeleon~\cite{xing2023pipeleon} shows that profile-guided, runtime program specialization can substantially improve P4 SmartNIC packet-processing performance on realistic workloads.

\paragraph{Characterization and limits.}
Recent studies emphasize that offload may shift, rather than eliminate, bottlenecks. For example, a holistic characterization of an off-path SmartNIC (BlueField-2) shows that communication paths and DMA/SoC constraints can dominate end-to-end benefits, and motivates designs that explicitly exploit multiple SmartNIC communication paths~\cite{wei2023offpath}. These results suggest that HO should be discussed together with \emph{observability} (how to measure offload effects), \emph{upgrade cadence} (hardware vs. software evolution), and \emph{failure domains} (what fails when the accelerator is saturated or misconfigured).

\paragraph{A recurring pattern: hybrid fast-path/slow-path partitioning.}
A common architectural response is to map throughput-critical and state/memory-intensive tasks onto different hardware tiers. Tiara~\cite{zeng2022tiara} exemplifies this approach for stateful L4 load
balancing via a three-tier pipeline that combines a programmable switch, FPGAs, and x86 servers, highlighting that \emph{heterogeneity} is often necessary for both scale and flexibility.

\begin{table}[t]
  \centering
  \small
  \begin{tabular}{p{3.5cm}p{5.8cm}p{5.8cm}}
    \toprule
    \textbf{Placement} & \textbf{What it buys} & \textbf{Typical constraints} \\
    \midrule
    Host user space & Fast iteration; rich libraries; flexible state &
    CPU overhead; jitter under load; NUMA/PCIe costs \\
    Host kernel / eBPF & Lower overhead; closer to NIC; easier enforcement &
    Verifier limits; complexity; constrained debugging \\
    SmartNIC / DPU & Offload host CPU; locality to NIC datapath; isolation knobs &
    Weaker cores; device-specific performance; limited memory hierarchy \\
    Switch/FPGA gateway & Line-rate fast path; low latency; shared enforcement point &
    Limited state; pipeline constraints; slower feature evolution \\
    \bottomrule
  \end{tabular}
  \caption{A practical HO view: \emph{where} a function runs is now a first-class design axis.}
  \label{tab:ho-placement}
\end{table}

\subsection{Cloud-native CNF era: additional design dimensions}
\label{sec:designspace-cloud-native}

The Kubernetes-first CNF era introduces several design dimensions that are less explicit in VM-centric NFV platforms. We summarize the most impactful ones below.

\textbf{Control model: workflow engines vs. reconciliation loops.}
Traditional MANO frameworks often encode lifecycle management as explicit workflows. In contrast, a Kubernetes-native approach expresses intent as a desired state and relies on continuous reconciliation loops implemented by operators/controllers~\cite{breitgand2021true_cloud_native_mano}.

\textbf{Day-2 operations as first-class concerns.}
Cloud-native deployments emphasize upgrades, rollbacks, configuration drift handling, and health remediation as routine operations. This favors platforms that can expose lifecycle hooks and safety mechanisms aligned with Kubernetes rollouts and operator logic~\cite{breitgand2021true_cloud_native_mano}.

\textbf{Cloud-deployability and reuse of commodity substrates.}
A key design goal is to minimize bespoke infrastructure while still meeting NFV requirements. Quadrant follows this principle by reusing Kubernetes and other cloud components and adding only targeted NFV-specific mechanisms when necessary (e.g., scheduling and packet-processing isolation)~\cite{wang2022quadrant}.

\textbf{Infrastructure profiles and conformance.}
Compared to ad-hoc NFVI stacks, the CNF era increases the need for shared infrastructure profiles and
validation criteria to reduce fragmentation across vendors and operators. The CNTT reference framework is a concrete attempt to standardize such a baseline for telecom cloud infrastructure~\cite{cntt_whitepaper}.

\begin{table}[t]
\centering
\footnotesize
\caption{CNF-era design dimensions that should be added to the NFV platform design space.}
\label{tab:cnf-era-designspace}
\begin{tabular}{p{3.2cm}p{5.2cm}p{6.2cm}}
\hline
\textbf{Dimension} & \textbf{Typical options} & \textbf{Implication to platform design} \\
\hline
Control model & Workflow engine vs. reconciliation loop & Impacts how lifecycle logic is expressed, verified, and evolved. \\
Day-2 ops & Imperative scripts vs. operator-managed rollouts & determines upgrade/rollback safety and operational automation. \\
Reuse of cloud substrate & Bespoke NFV stack vs. cloud-deployable reuse & Trades engineering effort against portability and ecosystem leverage. \\
Infrastructure profiles & Ad-hoc NFVI vs. standardized profiles & Affects portability across vendors and repeatability of validation. \\
\hline
\end{tabular}
\end{table}


\section{Open issues and challenges} \label{sec:challenge}
In this section, we outline future directions for NFV platform design and discuss the challenges associated with them.
Compared to the early NFV era (where the main focus was replacing dedicated appliances with software VNFs), recent progress has shifted the center of gravity toward (i) \emph{autonomous, closed-loop operations} driven by data and AI, (ii) \emph{end-to-end slice management} across RAN/core/transport/edge under explicit SLAs, and (iii) \emph{IoT/edge-native deployments} that stress scalability, intermittency, and energy constraints.
These trends are also increasingly reflected in operator-facing frameworks and standards that emphasize intent-driven automation, analytics functions, and digital twins as first-class capabilities~\cite{tmforum_autonomous_networks,etsi_zsm018_digital_twin,3gpp_slice_management,3gpp_ts23288_nwdaf}.

\subsection{AI in NFV}
Since 2020, ``AI in NFV'' has gradually shifted from using ML for isolated tasks (e.g., traffic prediction) to a broader goal of \emph{autonomous operations}, in which the platform continuously senses, analyzes, decides, and acts through closed loops.
Industry frameworks increasingly make this explicit: TM Forum defines autonomy maturity levels and promotes a transition toward highly autonomous networks, where intelligence is operationalized in production workflows rather than treated as an offline optimization tool~\cite{tmforum_autonomous_networks}.
In parallel, ETSI ZSM has been extending the zero-touch management framework toward \emph{intent-driven closed loops} and \emph{digital-twin integration} for safer what-if analysis and policy validation prior to actuation~\cite{etsi_zsm_overview,etsi_zsm018_digital_twin}.
On the 3GPP side, the 5G core introduces the Network Data Analytics Function (NWDAF) as a standardized analytics provider for network functions and OAM, with the specification evolving to cover more advanced analytics workflows and operational considerations (e.g., model-accuracy monitoring and distributed-learning features)~\cite{3gpp_ts23288_nwdaf}.

Despite this progress, several challenges remain.
First, the \emph{data problem} is still fundamental: NFV platforms must continuously collect, clean, and align multi-source telemetry (NF metrics, infrastructure counters, traces, and service-level indicators) while preserving privacy and access control.
This is particularly difficult in multi-vendor environments and in slices spanning multiple administrative domains, where the platform may only have partial observability and inconsistent semantics across data sources~\cite{3gpp_slice_management,3gpp_ts23288_nwdaf}.
Second, \emph{robustness and generalization} are hard to guarantee: models trained on historical behavior can fail under software upgrades, shifting traffic mixes, or rare failure modes; this elevates the importance of lifecycle management for models (validation, drift detection, rollback) as part of the platform rather than as ad-hoc engineering.
Third, \emph{safety and security of closed loops} becomes a first-class concern: once AI systems can trigger reconfiguration, scaling, or placement actions, the platform must prevent unsafe feedback cycles, enforce guardrails, and ensure that cross-loop interactions do not amplify incidents~\cite{etsi_zsm_overview}.
Finally, emerging ``agentic'' operational workflows (e.g., copilots/agents for troubleshooting and execution) raise additional concerns about accountability, reproducibility, and operator trust: platforms need mechanisms to record the evidence, decisions, and actions taken by AI components, especially for high-impact changes~\cite{tmforum_autonomous_networks}.

\subsection{Network slicing}
Network slicing has matured from a conceptual 5G enabler into an operationally demanding problem that stresses NFV platforms end-to-end.
3GPP slice management has evolved across releases: Rel-15 introduced the basics of slice lifecycle management; Rel-16 added SLA attributes and the notion of closed-loop automation; Rel-17 extended slicing to better support non-public networks and closed-loop assurance across multiple SLAs; and Rel-18 continues to explore more efficient provisioning and intent-driven slice management~\cite{3gpp_slice_management}.
The same line of work also expands the KPI framework to include slice-level performance evaluation and energy-efficiency KPIs, making sustainability and cost more explicit objectives of slice operations~\cite{3gpp_slice_management,3gpp_ts28554_kpi}.

However, delivering slices as \emph{predictable products} remains nontrivial.
First, slices are inherently \emph{cross-domain}: realizing an SLA requires coordinated control across RAN, core, transport, and increasingly the edge cloud, which may each be managed by different controllers and expose different actuation primitives.
The practical difficulty is not only deciding a per-domain configuration, but coordinating timescales and avoiding instability when multiple closed loops interact.
Second, slice isolation is multifaceted: it includes resource isolation (CPU, memory, I/O, NIC offloads), fault isolation, performance isolation under contention, and security isolation.
Recent work on non-public networks (NPNs) highlights concrete operational requirements (fault management, SLA monitoring, exposure to vertical customers, and stronger isolation expectations) that make these issues more pressing in enterprise settings~\cite{3gpp_npn_management}.
Security requirements also differ by vertical; 3GPP has continued to extend NPN-related security features across releases, reflecting the need for stronger and more customizable security postures in private deployments~\cite{3gpp_npn_security}.
Third, slice assurance increasingly relies on analytics pipelines: slice-level KPIs must be composed from constituent functions and infrastructure signals, which requires consistent measurement, data models, and often real-time streaming telemetry~\cite{3gpp_slice_management}.
Finally, open and disaggregated RAN ecosystems introduce new integration challenges: the O-RAN architecture explicitly supports intelligence-driven control via RIC components and standardized interfaces, but the resulting ecosystem demands careful engineering to ensure closed-loop slice assurance interoperability and robustness in multi-vendor deployments~\cite{oran_specs_updates_2022}.

\subsection{Integration with IoT}
IoT is no longer a single ``massive device connectivity'' scenario; it spans a continuum from ultra-low-power sensors to mid-tier broadband IoT devices and latency-sensitive industrial endpoints.
3GPP Release 17 introduced Reduced Capability (RedCap) NR devices, positioned between mMTC-style devices and full NR UEs, and Release 18 further refines this direction (e.g., enhanced RedCap) to broaden the practical IoT device ecosystem in 5G standalone deployments~\cite{3gpp_redcap}.
At the same time, private and non-public network deployments have become a major driver for industrial IoT, raising concrete requirements on manageability, exposure, and isolation for vertical customers~\cite{3gpp_npn_management,3gpp_npn_security}.
From the infrastructure perspective, edge computing has become a central complement to IoT: ETSI MEC has progressed toward a heterogeneous, multi-domain edge cloud, including federation, multi-tenancy/slicing considerations, intermittently connected components, and security enhancements~\cite{etsi_mec_overview}.

These developments create new challenges for NFV platforms.
First, IoT workloads stress \emph{scalability in control and data planes}: the platform must handle huge numbers of devices, frequent mobility/attachment events, and bursty signaling, while still meeting low-latency requirements for a subset of endpoints.
Second, IoT deployments are often \emph{geo-distributed and intermittently connected}: NFV platforms must reason about partial failures, degraded backhaul, and the placement of stateful functions close to devices, often under tight resource budgets at the edge~\cite{etsi_mec_overview}.
Third, energy becomes a first-order objective: IoT devices demand long battery life, while operators increasingly track energy efficiency at the slice and function levels, making energy-aware orchestration and scheduling more than an optimization ``nice-to-have''~\cite{3gpp_slice_management,3gpp_ts28554_kpi}.
Finally, IoT expands the attack surface: devices are deployed in less-controlled environments, supply-chain diversity is greater, and the platform must enforce strong isolation and authentication/authorization mechanisms consistent with private-network requirements~\cite{3gpp_npn_security}.
Overall, supporting IoT effectively will require NFV platforms that unify edge-aware orchestration, provide robust multi-tenant isolation, and enable energy-aware closed-loop operations, while remaining operable across heterogeneous hardware and evolving standards.

\section{Conclusion}\label{sec:conclusion}
As an emerging paradigm for shifting network management and service provisioning, NFV is expected to revolutionize next-generation telecommunication networks.
To accelerate the innovation and commercial adoption of NFV, a wide range of platforms has been implemented over the last six years. While sharing the ultimate objective of promoting NFV, they typically address distinct problems in the NFV ecosystem and adopt different design choices to achieve different performance metrics or service-layer agreements. Little work has been devoted to interpreting this large collection of platform implementations.
In this paper, we present a comprehensive survey of existing NFV platforms. After a brief review of typical NFV architectures, we present our taxonomy of existing NFV platforms by design purpose.
We then explore the design space and investigate the choices individual NFV platforms make to address different implementation challenges. We believe this work is comprehensive enough to serve as a first-hand guide for network operators, service providers, and developers to choose the most suitable NFV platforms or reinvent the wheel based on their specific requirements.

NFV has also evolved substantially beyond its early ``VM-centric'' form. In recent deployments, cloud-native principles have become increasingly influential: network functions are decomposed into microservices, managed declaratively, and operated through automated lifecycle pipelines, with Kubernetes widely used as the execution and management substrate for containerized network functions (CNFs). This shift expands the NFV platform scope from virtualized datapaths to a broader \emph{Telco Cloud} capability, where portability, upgrade cadence, and day-2 operations can be as critical as raw throughput. Consequently, platforms must make end-to-end choices that jointly account for orchestration semantics, software supply chains, and operational reliability.

At the same time, the NFV dataplane substrate is entering a heterogeneous era. High-speed packet I/O on commodity servers remains important, but platform designers increasingly leverage multiple execution targets, including in-kernel fast paths (e.g., eBPF/XDP), SmartNICs/DPUs, programmable switches, and FPGA/gateway offloads. Recent work suggests that acceleration is no longer merely a collection of isolated optimizations: it increasingly requires systematic partitioning of service chains across heterogeneous substrates, careful management of on-device resource contention, and toolchain support that reduces the engineering burden of generating accelerated variants. These trends reinforce one of the key takeaways of our design-space analysis: performance, portability, and evolvability are now coupled, and must be addressed as a coherent set of platform choices rather than independent knobs.

Beyond performance, operational properties have become first-class objectives. Modern platforms must support predictable multi-tenant behavior under consolidation, scale on multicore hardware without forcing intrusive rewrites of VNFs, and flexibly handle stateful services (including designs that reduce the coupling between state and compute). In addition, the community is placing renewed emphasis on diagnosability, accountability, and trustworthiness: as platforms become more programmable and open to third-party VNFs, operators increasingly need structured performance interfaces, auditable service chaining, and verification mechanisms that match real deployment artifacts. These directions complement classical NFV concerns (placement, scheduling, and migration) by strengthening the platform's ability to operate safely under frequent upgrades and complex failure modes.

Looking forward, we expect NFV platforms to be shaped by three tightly coupled forces. First, AI-driven closed-loop automation will increasingly shift from isolated analytics to intent-driven operations and continuous decision-making, raising new requirements for data quality, safety guardrails, and model lifecycle management. Second, network slicing will continue to stress NFV across domains (RAN/core/transport/edge), requiring robust cross-loop coordination, stronger isolation guarantees, and explicit KPI/SLA management for diverse tenants and verticals (including non-public networks). Third, IoT and edge deployments will push NFV platforms toward geo-distributed execution with stringent latency and energy constraints, motivating tighter integration with edge-computing ecosystems and more energy-aware orchestration. We hope that the taxonomy and design-space framework presented in this survey will help the community reason about these shifts, clarify the tradeoffs among competing platform designs, and accelerate the development of NFV platforms that are both performant and operable at scale.

\bibliographystyle{plain}   
\bibliography{biblio}

\end{document}